\documentclass[lettersize,journal,twoside]{IEEEtran}
\usepackage{amsmath,amsfonts,amssymb,amstext}
\usepackage{algorithmic}
\usepackage{algorithm}
\usepackage{array}
\ifCLASSOPTIONcompsoc
	\usepackage[caption=false,font=normalsize,labelfont=sf,textfont=sf]{subfig}
\else
	\usepackage[caption=false,font=footnotesize]{subfig}
\fi
\usepackage{textcomp}
\usepackage{stfloats}
\usepackage{url}
\usepackage{verbatim}
\usepackage{graphicx}
\usepackage{cite}
\usepackage{epsfig}
\usepackage{enumerate}
\usepackage{booktabs}
\usepackage{multirow}
\usepackage{array}

\newcolumntype{x}[1]{>{\centering\let\newline\\\arraybackslash\hspace{0pt}}p{#1}}
\newcolumntype{P}[1]{>{\centering\arraybackslash}p{#1}}
\newcolumntype{M}[1]{>{\centering\arraybackslash}m{#1}}

\usepackage{makecell}
\usepackage[table,xcdraw]{xcolor}

\newcommand{\splitcell}[1]{\begin{tabular}{@{}c@{}}#1\end{tabular}}
\newcommand{\bsplitcell}[1]{$\left[\splitcell{#1}\right]$}

\usepackage{tikz}

\begin{document}

\title{Task-specific Optimization of Virtual Channel Linear Prediction-based Speech Dereverberation Front-End for Far-Field Speaker Verification}

\author{Joon-Young Yang and Joon-Hyuk Chang,~\IEEEmembership{Senior Member,~IEEE}
}



\maketitle

\begin{abstract}
Developing a single-microphone speech denoising or dereverberation front-end for robust automatic speaker verification (ASV) in noisy far-field speaking scenarios is challenging.
To address this problem, we present a novel front-end design that involves a recently proposed extension of the weighted prediction error (WPE) speech dereverberation algorithm, the virtual acoustic channel expansion (VACE)-WPE.
It is demonstrated experimentally in this study that unlike the conventional WPE algorithm, the VACE-WPE can be explicitly trained to cancel out both late reverberation and background noise.
To build the front-end, the VACE-WPE is first independently (pre)trained to produce ``noisy" dereverberated signals.
Subsequently, given a pretrained speaker embedding model, the VACE-WPE is additionally fine-tuned within a task-specific optimization (TSO) framework, causing the speaker embedding extracted from the processed signal to be similar to that extracted from the ``noise-free" target signal.
Moreover, to extend the application of the proposed front-end to more general, unconstrained ``in-the-wild" ASV scenarios beyond controlled far-field conditions, we propose a distortion regularization method for the VACE-WPE within the TSO framework.
The effectiveness of the proposed approach is verified on both far-field and in-the-wild ASV benchmarks, demonstrating its superiority over fully neural front-ends and other TSO methods in various cases.
\end{abstract}

\begin{IEEEkeywords}
Speaker verification, speech dereverberation, weighted prediction error, deep neural network, single microphone, offline processing, virtual acoustic channel expansion.
\end{IEEEkeywords}

\section{Introduction} \label{sec1}
%
Automatic speaker verification (ASV) aims at recognizing if two speech segments are spoken by the same speaker.
Modern ASV systems typically involve transformation of utterances to fixed-dimensional vector representations, called speaker embeddings, which are required to encode speaker-discriminative information.
With the emergence of deep learning, deep speaker embedding (DSE) models, which exploit deep neural networks (DNNs) to build a speaker embedding extractor, have recently become conventional for ASV \cite{xvector18}.
However, although they are superior to the conventional Gaussian mixture model \cite{Reynolds_gmm_sv00} and i-vector \cite{ivector10} approaches, their performance is still limited in far-field speaking scenarios, in which speech recordings are severely degraded by reverberation and background noise \cite{voices_challenge_2019,ffsvc20,app_aware_bf19}.
Thus, several approaches have been proposed to build noise- and reverberation-robust ASV systems \cite{voices_challenge_2019,ffsvc20,app_aware_bf19,drvbf_ffsr18,assess_scse10,dfl_fe20,voiceidloss19,robust_sr_se_attn20,adv_bn_feats17,cgan_se_nr_sv17,mc_train_gplda12,mplda15,nplda20,fast_imap17,xmap20,adv_sep_adapt20,robust_sr_unsup_adv_inv20,selective_dse20,magneto20,mtadv_nr_emb19,adv_sv19,length_noise_aware_susr20,within_sample_var_sr20,multi_level_tl_ffsv20}, which can be broadly classified into three categories.

In the first category, the DSE model has a separate preprocessing front-end.
Although minimum variance distortionless response \cite{souden09pmwf} beamformers and the weighted prediction error (WPE) \cite{Nakatani-WPE-TASLP10,Yoshioka-MCLP-TASLP12} speech dereverberation algorithm are typically the most commonly selected in a multimicrophone setup \cite{app_aware_bf19,drvbf_ffsr18}, the former cannot be employed in a single-microphone system.
Particularly, in a single-microphone setup, a speech enhancement front-end may lead to performance degradation of ASV systems unless carefully designed or optimized \cite{assess_scse10,dfl_fe20,voiceidloss19}.
Therefore, several task-specific optimization (TSO) methods have been proposed to build DNN-based single-channel front-ends \cite{dfl_fe20,voiceidloss19,robust_sr_se_attn20}, wherein the DNNs are trained for acoustic feature enhancement \cite{dfl_fe20} or spectral masking \cite{voiceidloss19} by a task-specific method, by exploiting a pretrained DSE model.
In a similar context, in \cite{robust_sr_se_attn20}, a denoising DNN was jointly optimized with a pretrained DSE model.
Concurrently, the standard acoustic features were substituted by the bottleneck features learned from a DNN, which was adversarially trained to classify noise types \cite{adv_bn_feats17}.
Moreover, a conditional generative adversarial network \cite{Isola_cgan17} was used to build a speech enhancer for noise-robust ASV \cite{cgan_se_nr_sv17}.
%
%


In the second category, a separate postprocessing back-end operates in addition to the speaker embeddings.
The application of the standard probabilistic linear discriminant analysis (PLDA) \cite{plda07}, which may belong to this category, in noisy reverberant conditions was studied in \cite{mc_train_gplda12}.
More recently, mixture of PLDA \cite{mplda15} and neural PLDA \cite{nplda20} were proposed and shown to be effective as robust back-ends.
In \cite{fast_imap17}, an i-vector \cite{ivector10} denoising method based on statistical models was proposed and, subsequently, extended to x-vectors \cite{xvector18} using a denoising autoencoder \cite{xmap20}.
Concurrently, DNNs were employed to further disentangle speaker-relevant and -irrelevant factors from speaker embeddings \cite{adv_sep_adapt20,robust_sr_unsup_adv_inv20,selective_dse20}.
In \cite{magneto20}, a DNN was used to learn input-specific magnitude scalers for speaker embeddings, which, in turn, yielded nonmonotonically calibrated cosine similarity scores with improved discrimination ability.
%
%

The third category performs robust training of the DSE model.
In \cite{mtadv_nr_emb19,adv_sv19}, multi-task learning \cite{mtl_1997} and domain adversarial training \cite{dat_of_nn16} were adopted to build a DSE model invariant to different noise types.
In \cite{length_noise_aware_susr20,within_sample_var_sr20}, a DSE model used a clean utterance and its corrupted or cropped version as input, and the model was optimized to reduce the distance between the corresponding pair of embeddings.
In \cite{multi_level_tl_ffsv20} a far-field student DSE model was trained within the teacher--student learning \cite{knowledge_distill15} framework using a near-field teacher DSE model to learn a domain-invariant speaker embedding space.
%
%

%
%
%

In this study, we focus on the first category of ASV systems to deal with single-microphone far-field ASV problems.
Specifically, we present a novel front-end design based on the recently proposed virtual acoustic channel expansion (VACE)-WPE \cite{vace_wpe:is20,vace_wpe_taslp21}, which utilizes a dual-channel neural WPE \cite{neural_wpe:is17} in a single-microphone setup and a virtual signal generated by a DNN as the secondary input.
Similar to \cite{app_aware_bf19,dfl_fe20,voiceidloss19,robust_sr_se_attn20}, we apply a TSO framework such that given a pretrained VACE-WPE and a pretrained DSE model, the VACE-WPE is additionally fine-tuned to reduce the difference between a processed signal and a ``noise-free" target signal in the embedding space of the DSE model.
Herein, the pretraining of a front-end is conducted using ``noisy" but reverberation-free target signals, which is a key difference from the previous studies that applied a TSO framework \cite{dfl_fe20,voiceidloss19,robust_sr_se_attn20}.
In the subsequent TSO stage, a ``noise-free" target signal is employed based on the denoising capability of the VACE-WPE, a new property of the VACE-WPE demonstrated experimentally in this study.
%
Moreover, to prevent the potential degeneration of the front-end in non-far-field speaking conditions, we propose a distortion regularization method within the TSO framework by mimicking the response of the WPE algorithm to near-field speech recordings.
Experiments were conducted on two far-field ASV benchmarks \cite{voices_challenge_2019,ffsvc20} and one in-the-wild ASV benchmark \cite{sitw16} to further examine the generalizability of the front-ends to unconstrained conditions.
The main contributions of this article are summarized as follows:
\begin{itemize}
	\item We present experimental results that show that the VACE-WPE can be explicitly trained to cancel out noise components as well as late reverberations.
	\item We present a new front-end design that involves TSO of the VACE-WPE.
	The proposed TSO framework can be extended to train a fully neural front-end.
	\item We propose distortion-regularized (DR) TSO and present its effectiveness in various ASV scenarios.
\end{itemize}

\section{System Description} \label{sec2}

\subsection{Signal Model} \label{sec2:A}
Assume that $D$ microphones capture a source speech signal propagating in a reverberant enclosure.
In the short-time Fourier transform (STFT) domain, the signal observed from the $d$-th microphone can be approximated as follows \cite{Nakatani-WPE-TASLP10,Yoshioka-MCLP-TASLP12}:
\begin{align}
{Y}_{t,f,d} &= {X}_{t,f,d}+{N}_{t,f,d} = \sum_{\tau=0}^{l-1}{h_{\tau,f,d}^{*}}{S_{t-\tau,f}} + N_{t,f,d}\,,
\end{align}
where $Y_{t,f,d}$ is the observed signal; 
$X_{t,f,d}$ and $N_{t,f,d}$ are the speech and background noise components, respectively; 
and $S_{t,f}$ is the source speech.
$h_{t,f,d}$ denotes the room impulse response (RIR) from the source to the $d$-th microphone, whose duration is $l$, and $*$ denotes a complex conjugate operation.
The observed speech component, $X_{t,f,d}$, can be further decomposed into early arriving speech (i.\,e., the direct path plus early reflections) and late reverberation \cite{Nakatani-WPE-TASLP10} as follows:
\begin{align}
{Y}_{t,f,d} &= {X}_{t,f,d}^{\textrm{(early)}} + {X}_{t,f,d}^{\textrm{(late)}} + N_{t,f,d} = {Y}_{t,f,d}^{\textrm{(early)}} + {X}_{t,f,d}^{\textrm{(late)}} \\
&= \sum_{\tau=0}^{\Delta-1}{h_{\tau,f,d}^{*}}{S_{t-\tau,f}} + \sum_{\tau=\Delta}^{l-1}{h_{\tau,f,d}^{*}}{S_{t-\tau,f}} + N_{t,f,d}, \label{eqn:signal_model}
\end{align}
where $\Delta$ is the time frame index in the STFT domain and determines the duration of the RIR that contributes to the early arriving speech component.
Notably, the summation of ${X}_{t,f,d}^{\textrm{(early)}}$ and $N_{t,f,d}$ is represented as ${Y}_{t,f,d}^{\textrm{(early)}}$.

\subsection{Review of VACE-WPE} \label{sec2:B}

\subsubsection{Neural WPE} \label{sec2:B:1}
Under a noiseless assumption (i.\,e., ${N}_{t,f,d}=0$), the late reverberation component, ${X}_{t,f,d}^{\textrm{(late)}}$, can be approximated by the delayed linear prediction (LP) \cite{Nakatani-WPE-TASLP10} technique as follows:
\begin{align}
\hat{X}_{t,f,d}^{\textrm{(late)}} = \sum_{\tau=\Delta}^{\Delta+K-1}{\textbf{g}_{\tau,f,d}^{H}}{{\textbf{Y}}_{t-\tau,f}} =\, {\tilde{\textbf{g}}_{f,d}^{H}}{{\tilde{\textbf{Y}}}_{t-\Delta,f}}\,,
\end{align}
where ${\textbf{g}_{\tau,f,d}} \in \mathbb{C}^{D}$ represents the $K$-th order time-invariant LP filter coefficients for the output channel index, $d$; 
${{\textbf{Y}}_{t,f}} \in \mathbb{C}^{D}$ represents a $D$-channel stack of the microphone inputs; 
${\tilde{\textbf{g}}_{f,d}} \!=\! {[ {\textbf{g}_{\Delta,f,d}^{T}}, ..., {\textbf{g}_{\Delta+K-1,f,d}^{T}} ]}^{T} \!\in \mathbb{C}^{DK}$, ${\tilde{\textbf{Y}}_{t-\Delta,f}} \!=\! {[ {\textbf{Y}_{t-\Delta,f}^{T}}, ..., {\textbf{Y}_{t-(\Delta+K-1),f}^{T}} ]}^{T} \!\in \mathbb{C}^{DK}$; 
and $H$ and $T$ denote Hermitian and transpose operations, respectively.
Based on the delayed LP model, neural WPE \cite{neural_wpe:is17}, a variant of the iterative WPE \cite{Nakatani-WPE-TASLP10,Yoshioka-MCLP-TASLP12}, performs dereverberation as follows:
\begin{alignat}{3}
\textrm{Step 1)}
~~~~~~& {\lambda}_{t,f} = \frac{1}{D} \sum_{d}^{} {\lvert {\hat{Z}_{t,f,d}} \rvert}^2, && \label{eqn:nwpe_step1}\\
\textrm{Step 2)}
~~~~~~& \textbf{R}_{f} = \sum_{t}^{} \frac{{\tilde{\textbf{Y}}}_{t-\Delta,f} {\tilde{\textbf{Y}}_{t-\Delta,f}^{H}}} {{\lambda}_{t,f}} && \in \mathbb{C}^{DK \times DK}, \label{eqn:wpe_step2}\\
~~~~~~& \textbf{P}_{f} = \sum_{t}^{} \frac{{\tilde{\textbf{Y}}}_{t-\Delta,f} {{\textbf{Y}}_{t,f}^{H}}} {{\lambda}_{t,f}} && \in \mathbb{C}^{DK \times D}, \\
~~~~~~& \textbf{G}_{f} = {\textbf{R}_{f}^{-1}} {\textbf{P}_{f}} && \in \mathbb{C}^{DK \times D}, \label{eqn:wpe_step2-3} \\
\textrm{Step 3)}
~~~~~~& {{\textbf{Z}}_{t,f}} = {\textbf{Y}}_{t,f} - \textbf{G}_{f}^{H} {\tilde{\textbf{Y}}}_{t-\Delta,f}, && \label{eqn:wpe_step3}
\end{alignat}
where Eqs.\,(\ref{eqn:nwpe_step1})\,--\,(\ref{eqn:wpe_step3}) are derived under the maximum likelihood criterion based on a statistical signal model \cite{Nakatani-WPE-TASLP10,Yoshioka-MCLP-TASLP12}, $\textbf{G}_{f}$ is a matrix whose $d$-th column is ${\tilde{\textbf{g}}_{f,d}}$, and ${{\textbf{Z}}_{t,f}} =\hat{\textbf{Y}}_{t,f,d}^{\textrm{(early)}}$ is the $D$-channel stack of the dereverberated output signals.
In Eq.\,(\ref{eqn:nwpe_step1}), ${\lvert {\hat{Z}_{t,f,d}} \rvert}^2$ represents the power spectral density estimate of ${Z}_{t,f,d}$, and is obtained using the neural network operating in the log-scale power spectra (LPS) domain \cite{neural_wpe:is17} as follows:
\begin{align}
\ln{{\lvert {\hat{Z}_{t,f,d}} \rvert}^2} = \mathcal{F} \big( {\ln{{\lvert {{Y}_{d}} \rvert}^2}}; \Theta_\textrm{LPS} \big), \label{eqn:lpsnet}
\end{align}
where $\mathcal{F}(\,\cdot\,; \Theta_\textrm{LPS})$ represents the forward pass of the neural network, the supposed LPSNet, parameterized by $\Theta_\textrm{LPS}$.
The time-frequency (T--F) indices were omitted in $Y_{d}$ because neural networks typically consume multiple T--F units within a context as the input.
Note that because the WPE algorithm is not explicitly designed to remove background noise components \cite{Nakatani-WPE-TASLP10,Yoshioka-MCLP-TASLP12}, the LPSNet is trained to estimate the LPS of the ``noisy" early arriving speech (i.\,e., $\ln{{\lvert {Y}_{t,f,d}^{\textrm{(early)}} \rvert}^2}$) \cite{neural_wpe:is17}.

The LPSNet architecture employed in this study is identical to that described in \cite{vace_wpe_taslp21}, which comprises a bidirectional long short-term memory (BLSTM) layer with 400 units for each direction, followed by two fully connected layers with 800 exponential linear units (ELUs) \cite{elu_2015}.
The output layer is another fully connected layer with 513 linear units.
The LPSNet has $4.62$ million parameters.

\subsubsection{VACE-WPE} \label{sec2:B:3}
The VACE-WPE \cite{vace_wpe:is20,vace_wpe_taslp21} is neural WPE variant \cite{neural_wpe:is17} specifically designed to enhance the dereverberation performance of the WPE algorithm in a single-microphone offline processing scenario.
In addition to the LPSNet, the VACE-WPE exploits another neural network to generate a virtual secondary signal, for which a pair of actual and virtual signals is subsequently processed via the dual-channel neural WPE algorithm \cite{vace_wpe:is20,vace_wpe_taslp21}.
The neural network for the virtual signal generation, VACENet, is trained end-to-end 
such that the desired target signal is obtained from the actual output channel side of the dual-channel neural WPE.
Note that the LPSNet is trained in advance and not updated during the training of the VACENet.

\begin{figure}[t]
	\centering
	\includegraphics[width=3.15in]{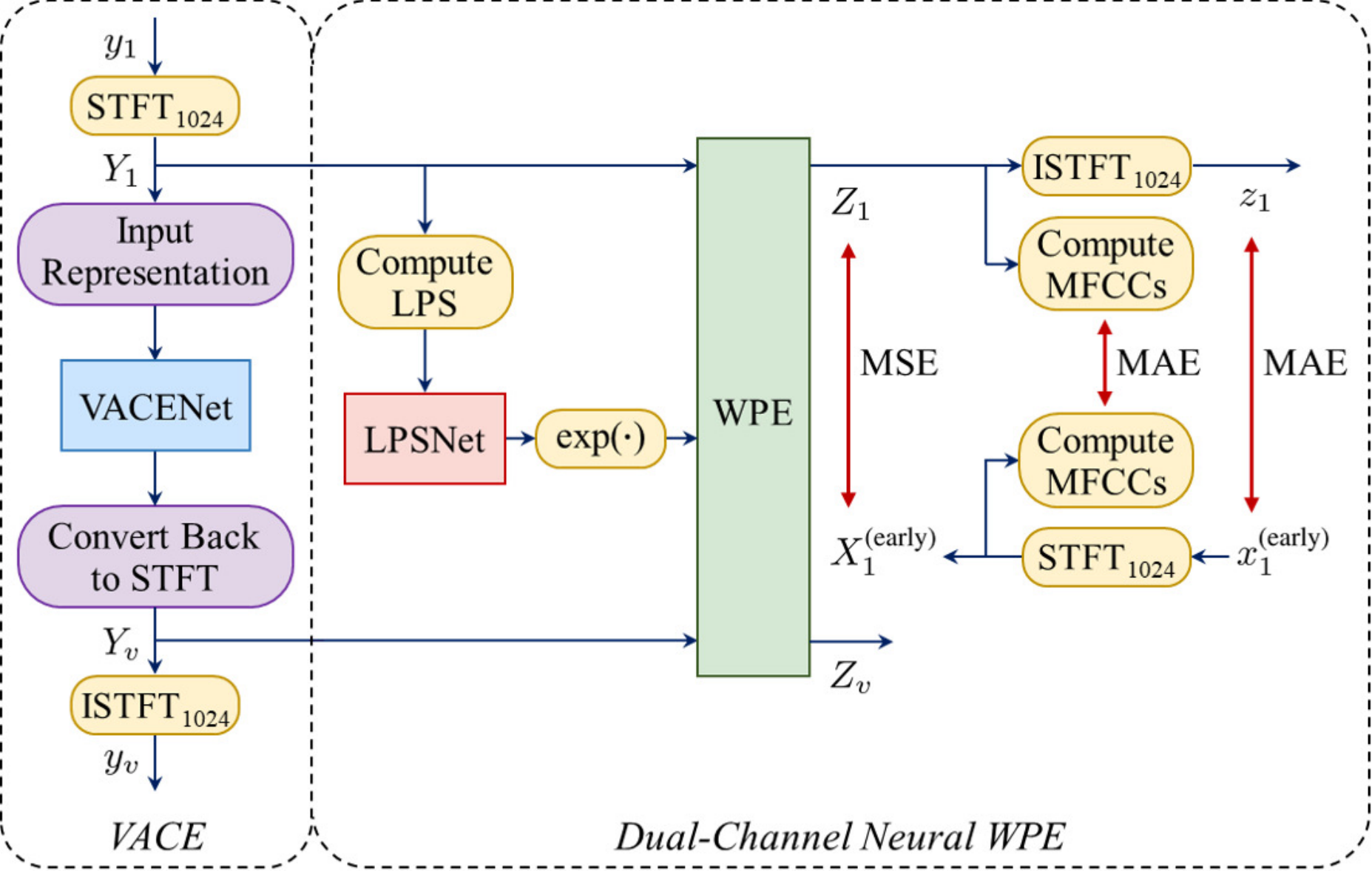}
	\caption{
		Block diagram of VACE-WPE \cite{vace_wpe_taslp21} front-end.
		Subscripts $1$ and $v$ denote actual and virtual channel signals, respectively.}
	\label{fig:vace_wpe}
\end{figure}

Fig.\,\ref{fig:vace_wpe} shows the block diagram of the VACE-WPE proposed in \cite{vace_wpe_taslp21}, where $Y_v$ and $y_v$ denote the virtual signal representations in the STFT and time domains, respectively.
Herein, the forward pass of the VACENet can be expressed as $Y_v = \mathcal{G} \left( Y_1; \Theta_\textrm{VACE} \right)$, and that of the VACE-WPE as $Z_1 = \mathcal{V} (Y_1; \Theta_\textrm{LPS},\Theta_\textrm{VACE})$, with $\Theta_\textrm{VACE}$ as the parameters of the VACENet.
Note that, as shown in Fig.\,\ref{fig:vace_wpe}, the VACE-WPE simplifies the power spectral density (PSD) estimation routine of Eq.\,(\ref{eqn:nwpe_step1}) to only exploit the observed actual signal and neglect the virtual channel counterpart \cite{vace_wpe_taslp21}.
%
\begin{align}
{\lambda}_{t,f} = {\lvert {\hat{Z}_{t,f,1}} \rvert}^2.
\label{eqn:nwpe_step1_simp}
\end{align}
Because of this modification, the generated virtual signal is independent of the PSD estimation routine in the forward pass of the VACE-WPE.
This promotes the role of the virtual signal in the LP filter estimation process of Eqs.\,(\ref{eqn:wpe_step2})\,--\,(\ref{eqn:wpe_step3}) \cite{vace_wpe_taslp21} (see Sections II-C6 and IV-B of \cite{vace_wpe_taslp21} for further details).

The entire procedure for training the VACENet is divided into pretraining and fine-tuning stages.
In the pretraining stage, the VACENet is trained to estimate the (noise-free) late reverberation component, $X_1^\textrm{(late)}$, as the output of the network, whereas in the fine-tuning stage, the VACENet is optimized to produce the noisy early arriving speech, $Y_1^\textrm{(early)}$, as the output of the simplified dual-channel neural WPE \cite{vace_wpe_taslp21}.

Two different types of loss functions are defined to train the VACENet \cite{vace_wpe_taslp21}.
\begin{align}
\begin{split}
{L}_\textrm{1}(A, B) &= \alpha\cdot [\textrm{MSE}(A^{r}, B^{r}) + \textrm{MSE}(A^{i}, B^{i})] \\
& + \beta\cdot \textrm{MSE}(\textrm{ln}|A|, \textrm{ln}|B|) + \gamma \cdot \textrm{MAE}(a, b), \label{eqn:L_freq}
\end{split} \\
{L}_\textrm{2}(A, B) &= {L}_\textrm{1}(A, B) + \eta \cdot \textrm{MAE}(A^\textrm{mfcc}, B^\textrm{mfcc}), \label{eqn:L_2}
\end{align}
where $a$, $A$, $\ln{|A|}$, and $A^\textrm{mfcc}$ represent the time-domain waveform, STFT coefficients, log-scale magnitude spectra, and mel-frequency cepstral coefficients, respectively; 
superscripts $r$ and $i$ denote the real and imaginary (RI) components of the STFT coefficients, respectively; 
$\alpha$, $\beta$, $\gamma$, and $\eta$ are scaling factors; 
and MSE$(\cdot,\cdot)$ and MAE$(\cdot,\cdot)$ are the mean squared error and mean absolute error in the inputs, respectively.
In the pretraining and fine-tuning stages, the VACENet is trained to respectively minimize the following loss functions \cite{vace_wpe_taslp21}:
\begin{align}
\begin{split}
{L}_\textrm{PT} &= {L}_\textrm{1}(\mathcal{G}(X_1; \Theta_\textrm{VACE}), X_1^\textrm{(late)}) \\
&+ {L}_\textrm{1}(\mathcal{G}(Y_1; \Theta_\textrm{VACE}), X_1^\textrm{(late)}), \label{eqn:L_pt} \\
\end{split} \\
\begin{split}
{L}_\textrm{FT} &= {L}_\textrm{2}(\mathcal{V}(X_1; \Theta_\textrm{LPS}, \Theta_\textrm{VACE}), X_1^\textrm{(early)}) \\
&+ {L}_\textrm{2}(\mathcal{V}(Y_1; \Theta_\textrm{LPS}, \Theta_\textrm{VACE}), Y_1^\textrm{(early)}). \label{eqn:L_ft}
\end{split}
\end{align}

The VACENet architecture employed in this study is identical to that described in \cite{vace_wpe_taslp21}.
Specifically, it is a U-Net \cite{unet2015} comprising a series of two-dimensional (2D) convolution (Conv2D) layers with kernel and stride sizes of $3\times3$ and $1\times1$, respectively, and gated linear unit \cite{glu:icml17} activations.
Exceptionally, Conv2D and transposed Conv2D layers with stride sizes of $2\times2$ are employed for downsampling and upsampling, respectively, with linear activations.
The output layer comprises a $1\times1$ Conv2D with a linear activation.
A key difference from the original U-Net \cite{unet2015} architecture is that our VACENet employs a dual-stream encoder to enhance the flexibility of the single-stream encoder \cite{vace_wpe_taslp21}.
For further details, please refer to \cite{vace_wpe_taslp21}.
The VACENet has $4.44$ million parameters.

\begin{table}[th]
	\caption{
		Detailed architecture of ResNet-34 DSE model. 
		Three-dimensional output feature maps are represented in 
		(channels $\times$ frequency $\times$ frames) format.
		$T$ denotes number of time frames, and the ``$\times$" symbol next to bracket indicates repetition of structure or multiple outputs
	}
	\label{tab:resnet34}
	\centering
	\begin{tabular}{M{1.4cm}|c|M{1.85cm}}
		\Xhline{3\arrayrulewidth}
		Name & Structure / Description & Output \\
		\Xhline{2\arrayrulewidth}
		Input & log-scale MFBEs & $1 \times 64 \times T$ \\ \hline
		SWMS  & sliding-window mean subtraction & $1 \times 64 \times T$ \\
		\Xhline{2.5\arrayrulewidth}
		Conv2D-0  & Conv2D: $3 \times 3$, 48, stride 1 & $48 \times 64 \times T$ \\ 
		\splitcell{ResBlock-1} & Conv2D: 
		\bsplitcell{$3 \times 3$, 48 \\ $3 \times 3$, 48 }\,$\times\,3$, stride 1 &
		$48 \times 64 \times T$ \\
		\splitcell{ResBlock-2} & Conv2D: 
		\bsplitcell{$3 \times 3$, 96 \\ $3 \times 3$, 96 }\,$\times\,4$, stride 2 &
		$96 \times 32 \times T/2$ \\
		\splitcell{ResBlock-3} & Conv2D: 
		\bsplitcell{$3 \times 3$, 192 \\ $3 \times 3$, 192 }\,$\times\,6$, stride 2 &
		$192 \times 16 \times T/4$ \\
		\splitcell{ResBlock-4} & Conv2D: 
		\bsplitcell{$3 \times 3$, 384 \\ $3 \times 3$, 384 }\,$\times\,3$, stride 2 &
		$384 \times 8 \times T/8$ \\
		\Xhline{2.5\arrayrulewidth}
		CDASP-0 & Dense: $\left[48 \times 16, 16 \times 48\right] \times 2$ & $\left[96\right] \times 2$ \\
		CDASP-1 & Dense: $\left[48 \times 16, 16 \times 48\right] \times 2$ & $\left[96\right] \times 2$ \\
		CDASP-2 & Dense: $\left[96 \times 32, 32 \times 96\right] \times 2$ & $\left[192\right] \times 2$ \\
		CDASP-3 & Dense: $\left[192 \times 64, 64 \times 192\right] \times 2$ & $\left[384\right] \times 2$ \\
		CDASP-4 & Dense: $\left[384 \times 128, 128 \times 384\right] \times 2$ & $\left[768\right] \times 2$ \\
		\Xhline{2.5\arrayrulewidth}
		Concat & concatenate all CDASP-$k$ outputs & $3$,$072$ \\ \hline
		Embedding  & Dense: $3$,$072 \times 256$,~\,BN: $\left[256\right] \times 2$ & $256$ \\ \hline
		AMSoftmax  & Dense: $256 \times 5$,$994$ & $5$,$994$ \\ 
		\Xhline{3\arrayrulewidth}
	\end{tabular}
\end{table}

\subsection{DSE Model} \label{sec2:C}
The architecture of the ResNet-34 \cite{resnet16} DSE model employed in this study is summarized in Table \ref{tab:resnet34}.
The input features are 64-dimensional (64D) log-scale mel-filterbank energies (MFBEs), which are normalized using the sliding-window mean subtraction (SWMS) method with a 3 s-long window.
Each ResBlock contains two stacks of Conv2D--batch normalization (BN) \cite{batchnorm}--rectified linear units (ReLUs) \cite{relu} and a residual connection; 
ResBlock-$k$ for $k\in\{2,3,4\}$ also contains a squeeze-and-excitation \cite{sqzext18} module with a reduction factor of $4$.
The speaker embedding is extracted as follows.
First, given that $U \in \mathbb{R}^{C \times F \times T}$ represents the output feature map of the Conv2D-0 layer or ResBlock-$k$ ($k\in\{1,2,3,4\}$), $U$ is reduced to $U^\text{mean} \in \mathbb{R}^{C \times T}$ and $U^\textrm{std} \in \mathbb{R}^{C \times T}$ by computing the mean and standard deviation along the frequency dimension, respectively.
Subsequently, both $U^\text{mean}$ and $U^\text{std}$ are subjected to a channel-dependent attentive statistics pooling (CDASP) \cite{ecapa_tdnn20}, which further reduces the time frame dimension.
\begin{gather}
e_{t,c}^{s} = {v_{c}^{s}}^{\top} \tanh(W^{s} U^{s} + b^{s}) + q_{c}^{s}, \nonumber \\ 
{\alpha}_{t,c}^{s} = \frac {\exp(e_{t,c}^{s})} {\sum_{\tau} \exp(e_{t,c}^{s})}, \nonumber \\ 
\tilde{\mu}_{c}^{s} = \sum_{\tau} {\alpha}_{t,c}^{s} U^{s}_{t,c}, ~~\tilde{\sigma}_{c}^{s} = \sqrt{\sum_{\tau} {\alpha}_{t,c}^{s} (U^{s}_{t,c})^2 - ({\tilde{\mu}_{c}^{s}})^2}, \nonumber
\end{gather}
where $s\in\{\textrm{mean, std}\}$, and $t$ and $c$ are the time frame and channel indices, respectively.
$W^{s} \in \mathbb{R}^{\frac{C}{r} \times C}$ and ${b}^{s} \in \mathbb{R}^{\frac{C}{r}}$ denote the projection matrix and bias vector, respectively.
${v}_{c}^{s}, \, {q}_{c}^{s} \in \mathbb{R}^{\frac{C}{r}}$ are channel-dependent weight and bias vectors for computing the attention scores, respectively, and $r$ is set as $3$ for all CDASP-$k$ ($k\in\{0,1,2,3,4\}$) layers.
Finally, all CDASP outputs are concatenated to a single vector and subsequently passed through the speaker embedding layer, which comprises an affine transform and a BN.
Note that we treat the output of the BN layer as the speaker embedding.
The ResNet-34 DSE model has $13.69$ million parameters up to the embedding layer.

The output layer comprises a weight matrix, whose column vectors are normalized to have unit lengths, and it projects the speaker embedding to the posterior probabilities of the identities of the training speakers.
Specifically, we adopt additive margin softmax (AMSoftmax) \cite{amsoftmax18} to define the cross-entropy loss for training the DSE model as follows:
\begin{equation}
L_\textrm{AM} = -\frac{1}{N} \sum_{i=1}^{N} \log \frac {e^{\lVert \mathbf{u}_{i} \rVert \cdot (\cos\theta_{y_{i}}-m)}} {e^{\lVert \mathbf{u}_{i} \rVert \cdot (\cos\theta_{y_{i}}-m)} + \sum_{j\neq y_{i}} e^{\lVert \mathbf{u}_{i} \rVert \cdot \cos\theta_{j}}}, \nonumber
\end{equation}
where $N$ is the mini-batch size, $i$ is a sample index, $\mathbf{u}_{i}$ and $\lVert \mathbf{u}_{i} \rVert$ denote the speaker embedding and its $\ell_{2}$-norm, respectively, ${y_{i}}$ denotes the target speaker index, and $m$ is the margin value, which is set as $0.2$ in this study.

\section{Proposed TSO Framework} \label{sec3}

\begin{figure}[t]
	\centering
	\includegraphics[width=3.36in]{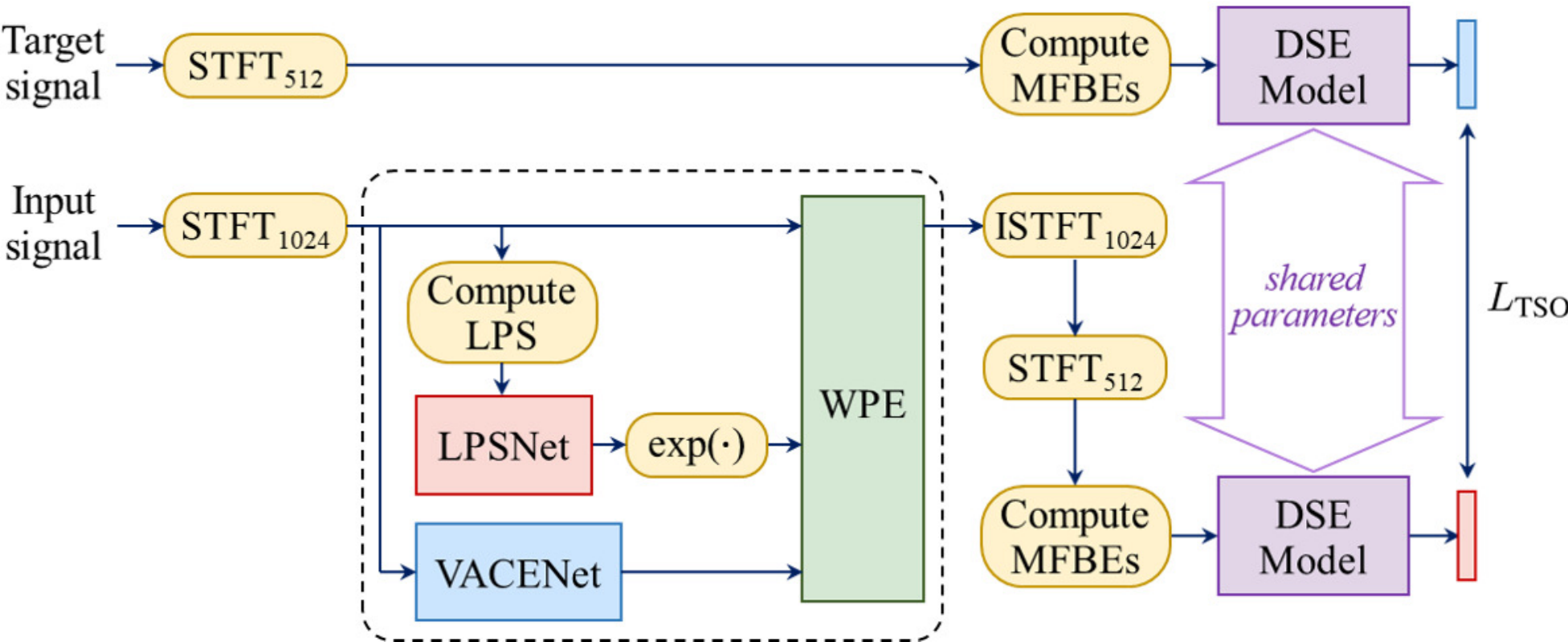}
	\caption{
		Schematic of proposed TSO framework for VACE-WPE \cite{vace_wpe_taslp21} front-end.
		Subscripts $1024$ and $512$ denote size of the fast Fourier transform, where corresponding (frame, hop) sizes are ($64$\,ms, $16$\,ms) and ($25$\,ms, $10$\,ms).
	}
	\label{fig:tso}
\end{figure}

\subsection{TSO of VACE-WPE} \label{sec3:A}
Fig.\,\ref{fig:tso} presents a schematic of the proposed TSO framework involving a pretrained DSE model and the VACE-WPE; the latter is also trained in advance as described in Section \ref{sec2:B:3}.
As shown in the figure, an input signal, possibly degraded by noise and reverberation, is processed by the VACE-WPE and subsequently transformed to the speaker embedding using the pretrained DSE model.
The target speaker embedding is directly extracted from the desired target signal.
Notably, different configurations are adopted for the VACE-WPE and the DSE model for the STFT analysis.
The window, hop, and fast Fourier transform (FFT) sizes are set as $64$\,ms, $16$\,ms, and $1024$ for the former and $25$\,ms, $10$\,ms, and $512$ for the latter, respectively \cite{app_aware_bf19}.
The loss function employed for the TSO is formulated as the negative cosine similarity between a pair of speaker embeddings as follows:
\begin{align}
\begin{split}
& {L}_\textrm{NCS}(A, B) \\
&= -\textrm{CS}\Big( \mathcal{H}(\mathcal{V}(A; \Theta_\textrm{LPS},\Theta_\textrm{VACE}); \Theta_\textrm{DSE}),\, \mathcal{H}(B; \Theta_\textrm{DSE}) \Big), \label{eqn:L_ncs} \\
\end{split}
\end{align}
where $\textrm{CS}(a, b) = {a^{\top}b} / ({\lVert{a}\rVert \cdot \lVert{b}\rVert})$, and $\mathcal{H}(\,\cdot\,; \Theta_\textrm{DSE})$ represents the forward pass of the DSE model up to the speaker embedding layer.
Finally, the TSO of the VACE-WPE is conducted by minimizing the following loss function:
\begin{align}
{L}_\textrm{TSO} = {L}_\textrm{NCS}(X_{1}, X_{1}^\textrm{(early)}) + {L}_\textrm{NCS}(Y_{1}, X_{1}^\textrm{(early)}). \label{eqn:L_tso}
\end{align}

It can be stated that the key feature of the proposed TSO method is the employment of a ``noise-free" early arriving speech, $X_{1}^\textrm{(early)}$, as the desired target signal while exploiting the VACE-WPE as the core algorithm of the front-end.
Specifically, because the conventional WPE algorithm was not designed to explicitly remove background noise \cite{Nakatani-WPE-TASLP10,Yoshioka-MCLP-TASLP12,neural_wpe:is17}, the ``noisy" early arriving speech, $Y_{1}^\textrm{(early)}$, was employed as the desired target signal for training the VACE-WPE in \cite{vace_wpe_taslp21}, as explained in Section \ref{sec2:B:3}.
In contrast, in this study, we employ a noise-free target signal during the TSO of the VACE-WPE; 
this is preceded by experimentally demonstrating that the VACE-WPE can be explicitly trained to remove noise components as well as late reverberations.
This denoising capability of the VACE-WPE is indeed a key factor supporting the proposed TSO framework, and is discussed in Section \ref{sec5:A}.

Another key difference from the other TSO methods \cite{dfl_fe20,voiceidloss19,robust_sr_se_attn20} is that instead of being trained from scratch or pretrained to estimate clean speech signals, the front-end is pretrained to dereverberate but not denoise prior to the TSO.

\begin{figure}[t]
	\centering
	\includegraphics[width=\linewidth]{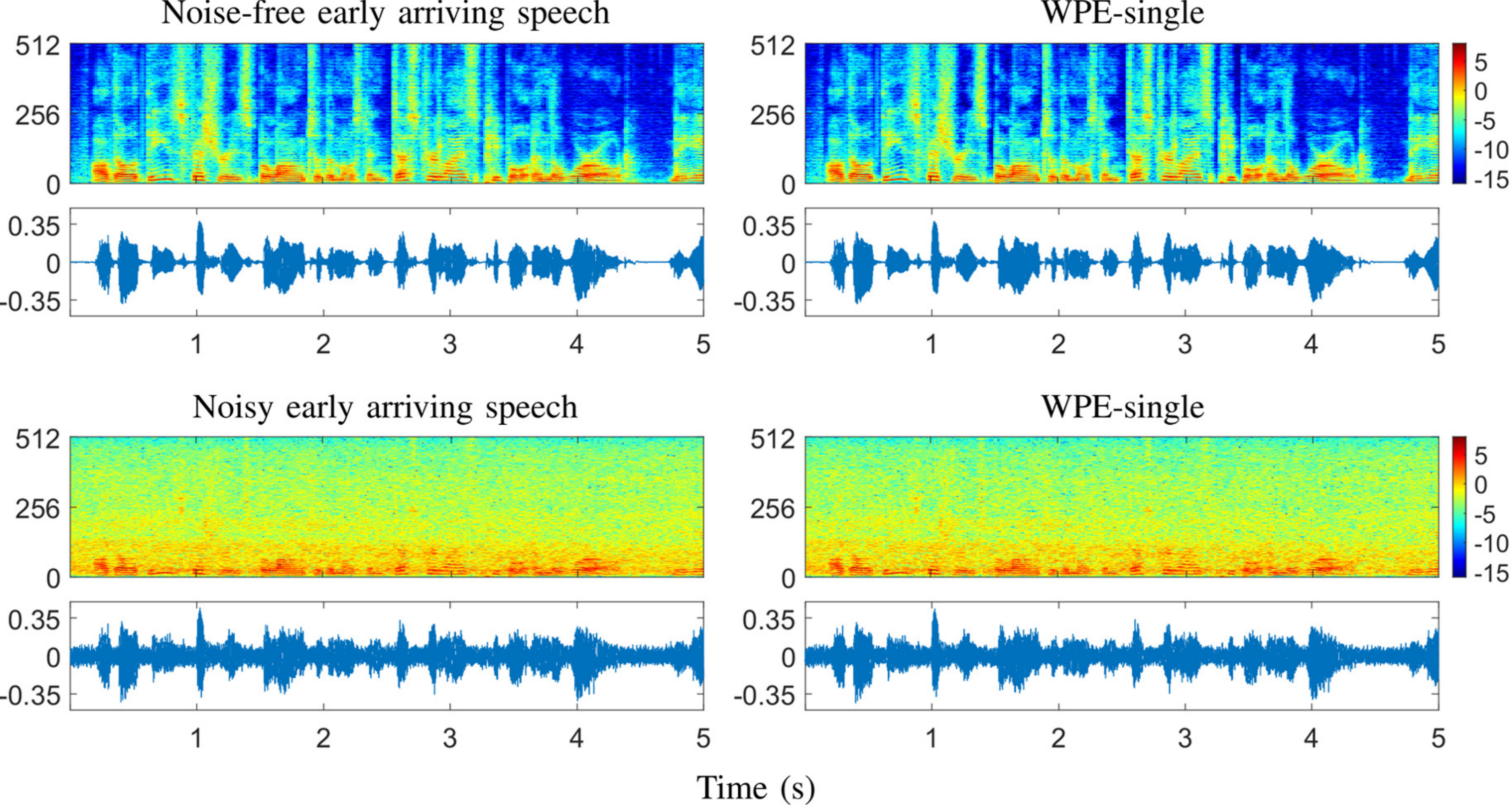}
	\caption{
		Dereverberation results of pretrained single-channel neural WPE given 
		noise-free (top) and noisy (bottom) early arriving speech signals.
	}
	\label{fig:nwpe_response}
\end{figure}

\subsection{DR-TSO of VACE-WPE} \label{sec3:B}
The primary role of a speech dereverberation or denoising front-end designed for far-field ASV is to effectively remove or suppress late reverberation and noise, without adversely affecting the speaker-discriminative components in the noisy reverberant input signal.
However, it would be more desirable if this front-end capability is not only limited to controlled far-field speaking scenarios, but extendable to more general, unconstrained conditions in which the utterances of speakers naturally contain different speaking distance, reverberation, and background noise levels.
To this end, we propose a method to regularize the TSO of the VACE-WPE to maintain non-far-field observations.
This is achieved by mimicking the operation regime of the WPE algorithm against near-field observations, which is detailed in the following.

Fig.\,\ref{fig:nwpe_response} shows the dereverberation regime of the single-channel neural WPE algorithm against noise-free and noisy early arriving speech signals $X_{1}^\textrm{(early)}$ and $Y_{1}^\textrm{(early)}$.
These speech samples were artificially generated using the signal model expressed in Eq.\,(\ref{eqn:signal_model}), with $\Delta$ representing the point $50$\,ms after the main peak of the RIR.
Because the input signals only contain early reflections and the WPE algorithm cannot explicitly remove noise components \cite{Nakatani-WPE-TASLP10,Yoshioka-MCLP-TASLP12,neural_wpe:is17}, the output signals are similar to their corresponding input signals.
Based on this simple observation, the distortion regularization term is proposed as follows:
\begin{align}
{L}_\textrm{DR} = {L}_\textrm{NCS}(X_{1}^\textrm{(early)}, X_{1}^\textrm{(early)}) + {L}_\textrm{NCS}(Y_{1}^\textrm{(early)}, Y_{1}^\textrm{(early)}), \label{eqn:L_dr}
\end{align}
which is a direct translation of the abovementioned dereverberation regime of the WPE, but defined in the learned embedding space of the pretrained DSE model.
Accordingly, a DR-TSO is conducted using the following loss function:
\begin{align}
{L}_\textrm{DR-TSO} &= {L}_\textrm{TSO} + {L}_\textrm{DR}. \label{eqn:L_dr_tso}
\end{align}

Notably, DR-TSO allows the desired dereverberation regime of the VACE-WPE against a noisy input signal to be interpreted as follows:
\begin{itemize}
	\item If the input signal is degraded by both reverberation and noise, the VACE-WPE must simultaneously perform dereverberation and denoising such that the speaker embedding extracted from the processed signal is similar to that extracted from the noise-free early arriving speech (Eq.\,(\ref{eqn:L_tso})).
	\item If the input signal is degraded by noise but is not severely reverberated, the VACE-WPE should operate such that the speaker embedding extracted from the processed signal is similar to that extracted from the raw undistorted input signal (Eq.\,(\ref{eqn:L_dr})).
\end{itemize}

\section{Experimental Setup} \label{sec4}


\subsection{ASV Benchmark Datasets} \label{sec4:A}
Three different ASV benchmarks Voices Obscured in Complex Environmental Settings (VOiCES) Challenge 2019 \cite{voices_challenge_2019}, Speakers in the Wild (SITW) \cite{sitw16}, and Far-field Speaker Verification Challenge (FFSVC) 2020 \cite{ffsvc20} are utilized to validate the proposed approach.

The VOiCES Challenge 2019 dataset \cite{voices_challenge_2019} contains noisy far-field English speech recordings, which are a re-recorded subset of the LibriSpeech \cite{librispeech} corpus.
More than 12 different combinations of microphone types and locations as well as 3 different rooms and 4 noise types (e.\,g., room ambient, babble, music, and television), are included \cite{voices_challenge_2019}.

The SITW dataset \cite{sitw16} comprises speech utterances recorded in unconstrained ``in-the-wild" conditions (e.\,g., interviews, conversational dialogues, and monologues), and thus, naturally contains noise, reverberation, and compression artifacts.
In this study, we only consider the ``core" segments \cite{sitw16}, which do not contain speech overlaps of multiple speakers.

The FFSVC 2020 dataset \cite{ffsvc20} contains Chinese Mandarin utterances recorded with a cellular phone and six four-channel circular microphone arrays at a distance.
The speaker enrollment and tests are conducted using the cellular phone and one of the microphone array recordings, respectively.
We adopt ``Task 2" \cite{ffsvc20}, which is a text-independent ASV task using a single microphone array, and only use the first channel recordings to compute the ASV scores.
The cellular phone recordings are downsampled to 16 kHz.

\renewcommand{\arraystretch}{1.1}
\begin{table}[]
	\caption{Statistics of ASV Benchmark Datasets}
	\label{tab:db_statistics}
	\centering
	\begin{tabular}{c|c|c|c|c|c|c}
		\Xhline{2\arrayrulewidth}
		Dataset & part & \#spk & \#utt & dur & \#tar & \#non   \\ \Xhline{2\arrayrulewidth}
		\multirow{2}{*}{\begin{tabular}[c]{@{}c@{}}VOiCES\\ 2019 \cite{voices_challenge_2019}\end{tabular}} & dev & 196 & 15,902 & 15.7\,s & 20,096 & 3,985,792  \\ \cline{2-7} 
		& eval & 100 & 11,392 & 15.8\,s & 36,443 & 3,571,073  \\ \Xhline{2\arrayrulewidth}
		\multirow{2}{*}{SITW \cite{sitw16}} & dev & 119 & 825 & 37.1\,s & 2,597 & 335,629  \\ \cline{2-7} 
		& eval & 180 & 1,202  & 37.4\,s & 3,658  & 718,130  \\ \Xhline{2\arrayrulewidth}
		\multirow{2}{*}{\begin{tabular}[c]{@{}c@{}}FFSVC\\ 2020  \cite{ffsvc20}\end{tabular}}  & dev & 104 & 29,436 & 3.1\,s  & 4,909  & 48,915  \\ \cline{2-7} 
		& eval & N/A & 83,130 & 3.1\,s  & 12,000 & 1,200,000 \\ 
		\Xhline{2\arrayrulewidth}
	\end{tabular}
\end{table}

The statistics of the three datasets are summarized in Table \ref{tab:db_statistics}, where the dataset partition, number of speakers and utterances, average duration in seconds, and number of target and nontarget trials are provided from the second to the last column, in that order.

\subsection{Training Datasets} \label{sec4:B}
\subsubsection{Training Dataset for DSE Model} \label{sec4:B:1}
The DSE model was trained using the ``dev" part of the VoxCeleb2 \cite{voxceleb2} dataset, which contains 1,092,009 utterances from 5,994 speakers.
Four additional copies were created for data augmentation purposes, whereby the simulated RIRs described in \cite{kaldi_rirs}, audio clips from the Deep Noise Suppression (DNS) Challenge 2020 dataset \cite{dns_challenge_2020}, music recordings from the MUSAN \cite{musan_corpus} dataset, and babble noise simulated using the ``us-gov" part of the MUSAN speech \cite{musan_corpus} corpus were used to corrupt the original set of utterances.

\subsubsection{Training Dataset for VACE-WPE} \label{sec4:B:2}
The modified LibriSpeech-80h dataset, which is provided as the training dataset for the automatic speech recognition task of the VOiCES Challenge 2019 \cite{voices_challenge_2019}, was used as the clean speech corpus for training the LPSNet and the VACENet.
First, we filtered out the speakers and utterances that overlap with those included in the ASV trial sets of the VOiCES Challenge 2019 dataset \cite{voices_challenge_2019}.
Subsequently, we employed an energy-based voice activity detector (VAD) to trim the silent regions and further excluded the utterances of duration less than 2.8 s \cite{vace_wpe_taslp21}.
Consequently, we obtained 14,985 utterances from 184 speakers, with an average duration of 12.3 s.

Note that the mismatch in training datasets for the front-end and the DSE model involves a system development scenario in which the DSE model is fully accessible for TSO, but the dataset on which it was trained is unknown or known yet unavailable.
This scenario is typical for developing a separate front-end (as pointed out in \cite{cam:icassp21}) and requires a front-end design that generalizes to a training data mismatch.

\subsection{Training Specifications} \label{sec4:C}
\subsubsection{Training of DSE Model} \label{sec4:C:1}
A single mini-batch was formed by collecting 32 randomly cropped chunks of 64D log-MFBEs, whose size was fixed as 250 frames; a VAD was not applied.
A single training epoch was defined as the iterations over 30,000 mini-batches.
We mostly followed the training scheme and hyperparameter setup described in \cite{large_margin_softmax_sv19}.
The DSE model was trained using the stochastic gradient descent optimizer with an initial learning rate of 0.01, which was halved every time the validation loss did not improve in three consecutive epochs.
The training was stopped if the validation loss plateaued for eight consecutive epochs.
The weights of the model were subject to $\ell_{2}$-regularization with a scale of 0.01 \cite{large_margin_softmax_sv19}, and a dropout \cite{dropout} was applied for every third mini-batch with a rate of 0.2.
The logit annealing strategy, which gradually applies AMSoftmax from the early stage of the training, was adopted to stabilize the training \cite{large_margin_softmax_sv19}.

To achieve ASV performances competitive to those reported in \cite{ffsvc20,great_ffsvc20,stc_ffsvc20} on the FFSVC 2020 benchmark, we additionally fine-tuned the pretrained DSE model using the ``train" portion of the FFSVC 2020 dataset.
We concatenated the recordings and obtained 4,920 utterances from 120 speakers.
The fine-tuning was conducted by first retraining only the output layer from scratch and subsequently training the entire model, as described in \cite{vace_wpe_taslp21}.
This additionally fine-tuned DSE model was only used to produce the ASV scores for the FFSVC 2020 benchmark and not to perform the TSO.

\subsubsection{Training of VACE-WPE} \label{sec4:C:2}
The training of the LPSNet and the VACENet was conducted as described in \cite{vace_wpe_taslp21}.
The training samples were generated on-the-fly using noise segments and RIRs randomly sampled from the DNS Challenge 2020 \cite{dns_challenge_2020} and simulated RIR \cite{kaldi_rirs} datasets, respectively.
The duration and the signal-to-noise ratio (SNR) were randomly chosen between 2.4 s and 2.8 s and between 3 dB and 20 dB, respectively.
We set $\alpha=1$, $\beta=0.04$, and $\gamma=5$ in Eq.\,(\ref{eqn:L_pt}) and $\alpha=1$, $\beta=0.1$, $\gamma=5$, and $\delta=0.2$ in Eq.\,(\ref{eqn:L_ft}) \cite{vace_wpe_taslp21}.
In the fine-tuning stage of the VACENet, the LP delay, $\Delta$, was fixed as $3$, and the tap size, $K$, was randomly selected within set $S_{K} = \{ K \,|\, K_{l} \!\leq K \!\leq K_{u} \} \subset \mathbb{Z}^\mathbf{+}$ for every single mini-batch.
Moreover, $K_l$ was fixed as $4$, and $K_u$ was initially set as $6$ and gradually increased to $21$ as the training progressed \cite{vace_wpe_taslp21}.
The Adam optimizer \cite{adam_optimizer} was employed.
The weights were $\ell_{2}$-regularized with a scale of $10^{-5}$, and gradient clipping \cite{gradient_clipping} with a global norm threshold of $3.0$ was applied.
For further details, please refer to Sections III-D, III-E, and IV-F of \cite{vace_wpe_taslp21}.

In the TSO stage, the parameters of the LPSNet and the DSE model were frozen, and only the VACENet was optimized using either the loss function in Eq.\,(\ref{eqn:L_tso}) or Eq.\,(\ref{eqn:L_dr_tso}).
A training epoch was defined as the iterations over 9,000 mini-batches, and the training was conducted for 30 epochs using the Adam optimizer \cite{adam_optimizer}.
The learning rate was initially set as $10^{-5}$ and annealed by $0.3$ after the 23rd epoch.
For every mini-batch, $K$ was randomly chosen within set $S_{K}$, where $K_{l}$ was fixed as $4$, and $K_{u}$ was initially set as $6$ and subsequently increased to $9$, $12$, $15$, $18$, and $21$ after the 5th, 9th, 13th, 16th, and 19th epochs, respectively.

\subsection{Evaluation Methods and Metrics} \label{sec4:D}
\subsubsection{ASV Scoring} \label{sec4:D:1}
Given a pair of enrollment and test segments, the ASV score was obtained as the cosine similarity between the corresponding pair of speaker embeddings.
To transparently observe the impact of adopting a front-end on the ASV performance, techniques such as global mean subtraction, PLDA \cite{plda07}, and score normalization \cite{asnorm:icassp11} were not applied.
Score calibration was also not considered; 
rather, we focused on analyzing the effect of adopting different front-ends on the discrimination ability of the ASV system.

For the single-channel WPE and VACE-WPE front-ends, the tap sizes of the LP filter were set as $K=30$ and $K=15$, respectively, for all ASV benchmarks.

\subsubsection{Evaluation Metrics} \label{sec4:D:2}
The minimum detection cost function (minDCF) \cite{nist_sre10_eval_plan} and the equal error rate (EER) were employed as the evaluation metrics.
The minDCF was computed with the prior probability of target speaker occurrence ($P_\textrm{tar}$) set as $0.01$ and the costs for both missed ($C_\textrm{miss}$) and false ($C_\textrm{fa}$) alarms set as $1$; 
it was the common primary metric employed by the three benchmarks \cite{voices_challenge_2019,ffsvc20,sitw16}.
The EER evaluates the operating point at which the missed and false alarm rates are identical.

\section{Experimental Results and Analysis} \label{sec5}

\subsection{Denoising Capability of VACE-WPE} \label{sec5:A}
\begin{figure}[t]
	\centering
	\includegraphics[width=\linewidth]{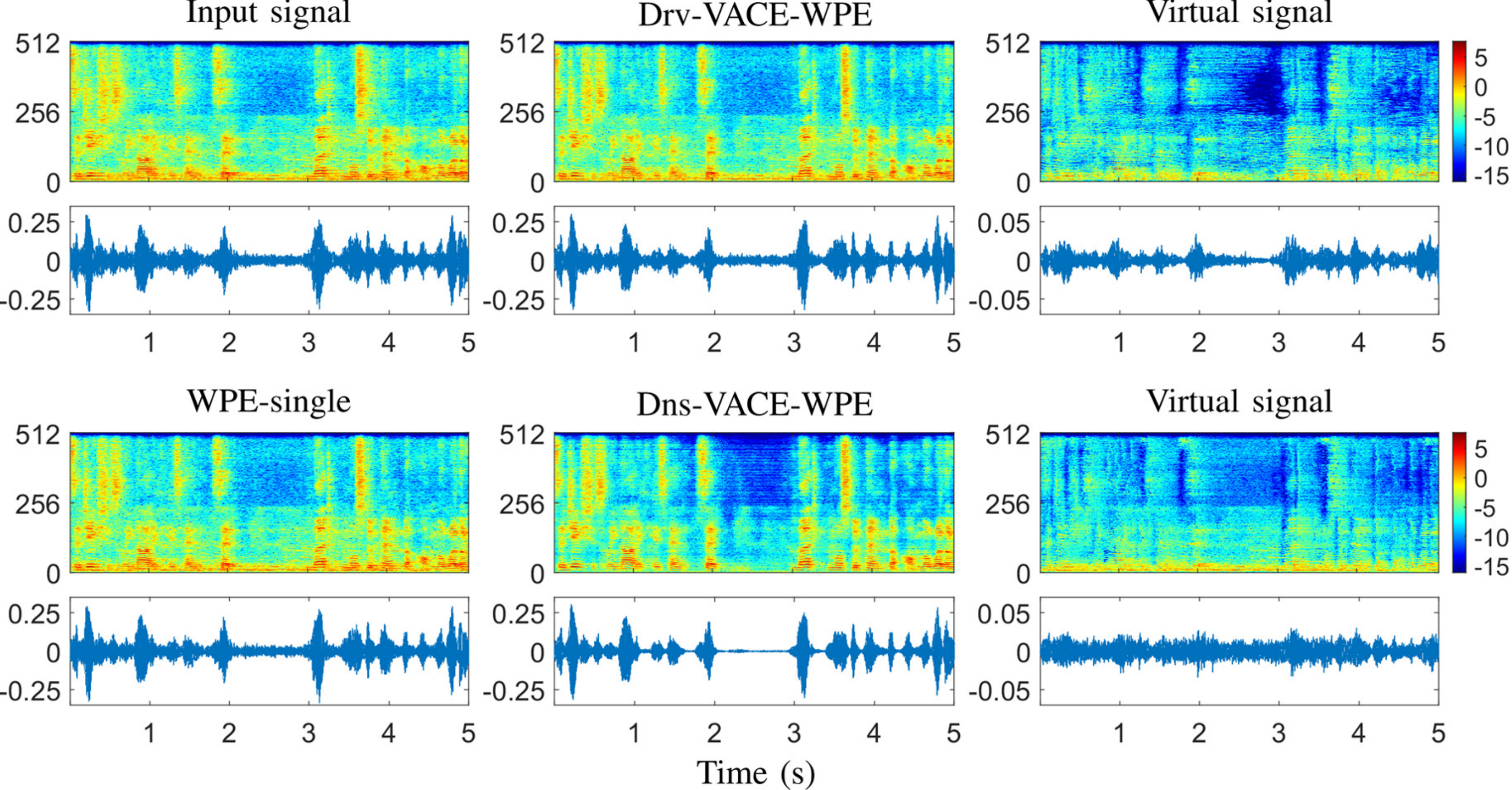}
	\caption{
		Comparison of output signals obtained using WPE-single, Drv-VACE-WPE, and Dns-VACE-WPE.
		Virtual signals on top and bottom were obtained from Drv-VACE-WPE and Dns-VACE-WPE, respectively.
	}
	\label{fig:comp_dns_drv}
\end{figure}

To first examine the explicit denoising capability of the VACE-WPE, in the fine-tuning stage of the VACENet, we simply substituted $Y_{1}^\textrm{(early)}$ in the second term of Eq.\,(\ref{eqn:L_ft}) with $X_{1}^\textrm{(early)}$; 
all the other configurations were maintained and TSO was not performed.
Fig.\,\ref{fig:comp_dns_drv} shows the output signals of the single-channel WPE (or WPE-single) and the two VACE-WPE variants given a noisy reverberant speech segment taken from the VOiCES Challenge 2019 ``eval" dataset.
Prefixes ``Drv-" and ``Dns-" denote the VACE-WPE front-ends trained using noisy and noise-free target signals, respectively.
Noticeably, the Dns-VACE-WPE is significantly better at cancelling out noise components than the WPE-single or Drv-VACE-WPE.
This is interesting because Drv-VACE-WPE and Dns-VACE-WPE as well as the WPE-single share the same LPSNet, which was trained to estimate the LPS of the ``noisy" early arriving speech.
We believe that this denoising capability of the VACE-WPE is enabled by the algorithmic flexibility introduced by the virtual signal within the end-to-end optimization procedure.
Specifically, when the VACE-WPE is trained end-to-end to output the desired target signals, the VACENet is optimized to produce a virtual signal that enables the desired operation of the VACE-WPE on the set of equations in Eqs.\,(\ref{eqn:nwpe_step1_simp}),\,(\ref{eqn:wpe_step2})--(\ref{eqn:wpe_step3}).

The virtual signals obtained from the Drv-VACE-WPE and the Dns-VACE-WPE are exhibited in the rightmost panel of Fig.\,\ref{fig:comp_dns_drv}, demonstrating marked differences in the required operation regime of the VACE-WPE.
Specifically, compared to the virtual signal of the former, that of the latter seems to contain more noise.

Table \ref{tab:resnet34_wpe} provides the average SNRs and the non-intrusive normalized signal-to-reverberation modulation energy ratios (SRMRs) \cite{non_intrusive_srmr} of the target ASV benchmark datasets.
The SRMR captures the higher-frequency temporal envelope characteristics of reverberant speech and does not require a reference signal for the metric computation; 
a high SRMR implies a less reverberant and less noisy signal.
The waveform amplitude distribution analysis \cite{wada_snr08} method was employed to estimate the SNR of the real speech recordings.
Noticeably, the output signals of the Dns-VACE-WPE yielded SNRs and SRMRs significantly higher than those of the unprocessed signals, WPE-single, and Drv-VACE-WPE on all three ASV benchmark datasets.

As explained briefly in Section \ref{sec3:A}, based on this denoising capability of the VACE-WPE, we designed the proposed TSO framework to 
consider 
a noise-free target signal.
This is noteworthy because conventionally, the WPE algorithm cannot perform explicit denoising \cite{Nakatani-WPE-TASLP10,Yoshioka-MCLP-TASLP12}.

\renewcommand{\arraystretch}{1.05}
\begin{table*}[]
	\caption{
		ASV performance on VOiCES Challenge 2019, SITW core--core, and FFSVC 2020 Task 2 (ResNet-34, WPE-based front-ends)
	}
	\label{tab:resnet34_wpe}
	
	\centering
	\begin{tabular}{@{}|M{2.7cm}|M{2.3cm}|M{1.5cm}M{1.5cm}|M{1.5cm}M{1.5cm}|M{1.5cm}M{1.5cm}|@{}}
		\Xhline{3\arrayrulewidth}
		\multirow{2}{*}{Method} &
		\multirow{2}{*}{Metric} &
		\multicolumn{2}{c|}{VOiCES 2019 \cite{voices_challenge_2019}} &
		\multicolumn{2}{c|}{SITW \cite{sitw16}} &
		\multicolumn{2}{c|}{FFSVC 2020 \cite{ffsvc20}} \\ \cline{3-8}
		&         & dev   & eval   & dev   & eval  & dev   & eval  \\ 
		\Xhline{2.6\arrayrulewidth}
		\multirow{2}{*}{Unprocessed}
		& minDCF~/~EER(\%)  &0.116~/~1.08 &0.360~/~5.11 &0.135~/~1.33 &0.151~/~1.51 &0.575~/~4.53 &0.577~/~5.63 \\ 
		& ~SRMR~/~SNR(dB)   &~2.15~/~11.8  &1.97~/~8.9  &~2.89~/~20.4  &~2.90~/~19.6  &~2.00~/~15.4  &~1.97~/~13.5  \\ 
		\Xhline{2.6\arrayrulewidth}
		\multirow{2}{*}{WPE-single}
		& minDCF~/~EER(\%)  &0.102~/~1.01 &0.321~/~4.68 &\textbf{0.133}~/~\textbf{1.31} &\textbf{0.146}~/~\textbf{1.49} &0.568~/~4.48 &0.569~/~5.68 \\ 
		& ~SRMR~/~SNR(dB)   &~2.54~/~14.4  &~2.33~/~11.1  &~2.95~/~21.3  &~2.96~/~20.4  &~2.45~/~16.2  &~2.40~/~13.6  \\ 
		\Xhline{1.0\arrayrulewidth}
		\multirow{2}{*}{TSO$_\mathcal{C}$-WPE-single}
		& minDCF~/~EER(\%)  &0.102~/~1.02 &0.320~/~4.70 &0.132~/~1.33 &0.148~/~1.51 &0.585~/~4.59 &0.571~/~5.82 \\ 
		& ~SRMR~/~SNR(dB)   &~2.52~/~14.3  &~2.32~/~11.1  &~2.95~/~21.2  &~2.96~/~20.3  &~2.43~/~16.6  &~2.38~/~13.8  \\ 
		\Xhline{2.6\arrayrulewidth}
		\multirow{2}{*}{Dns-VACE-WPE}
		& minDCF~/~EER(\%)  &0.113~/~1.12 &0.438~/~5.60 &0.199~/~2.24 &0.245~/~2.31 &0.707~/~6.52 &0.760~/~8.41 \\ 
		& ~SRMR~/~SNR(dB)   &~\underline{3.10}~/~\underline{29.2}  &~\underline{3.16}~/~\underline{24.2}  &~\underline{3.31}~/~\underline{32.8}  &~\underline{3.36}~/~\underline{32.9}  &~\underline{3.20}~/~\underline{28.1}  &~\underline{3.17}~/~\underline{27.4}  \\ 
		\Xhline{1.0\arrayrulewidth}
		\multirow{2}{*}{Drv-VACE-WPE}
		& minDCF~/~EER(\%)  &0.099~/~0.99 &0.331~/~4.71 &\textbf{0.131}~/~\textbf{1.31} &0.148~/~1.52 &0.591~/~4.58 &0.585~/~5.92 \\ 
		& ~SRMR~/~SNR(dB)   &~2.87~/~16.3  &~2.73~/~13.0  &~2.99~/~21.6  &~3.00~/~21.1  &~2.96~/~17.0  &~2.88~/~14.2  \\ 
		\Xhline{2.6\arrayrulewidth}
		\multirow{2}{*}{TSO$_\mathcal{N}$-VACE-WPE}
		& minDCF~/~EER(\%)  &0.096~/~0.97 &0.309~/~4.56 &0.132~/~1.32 &0.149~/~1.49 &0.570~/~4.36 &0.561~/~5.63 \\ 
		& ~SRMR~/~SNR(dB)   &~2.80~/~15.9  &~2.63~/~12.4  &~2.97~/~21.8  &~2.99~/~20.9  &~2.82~/~17.2  &~2.75~/~14.4  \\ 
		\Xhline{1.0\arrayrulewidth}
		\multirow{2}{*}{TSO$_\mathcal{C}$-VACE-WPE}
		& minDCF~/~EER(\%)  &\textbf{0.091}~/~\textbf{0.96} &\textbf{0.250}~/~\textbf{3.67} &0.133~/~1.41 &0.155~/~1.66 &0.556~/~4.37 &0.563~/~5.57 \\ 
		& ~SRMR~/~SNR(dB)   &~2.90~/~22.4  &~2.82~/~17.3  &~3.05~/~25.4  &~3.08~/~25.0  &~2.93~/~21.5  &~2.88~/~18.5  \\ 
		\Xhline{1.0\arrayrulewidth}
		\multirow{2}{*}{DR-TSO$_\mathcal{C}$-VACE-WPE}
		& minDCF~/~EER(\%)  &\textbf{0.091}~/~\textbf{0.97} &\textbf{0.263}~/~\textbf{3.96} &\textbf{0.131}~/~1.35 &0.150~/~\textbf{1.46} &\textbf{0.549}~/~\textbf{4.33} &\textbf{0.556}~/~\textbf{5.49} \\ 
		& ~SRMR~/~SNR(dB)   &~2.89~/~22.2  &~2.80~/~17.3  &~2.98~/~22.6  &~3.00~/~22.1  &~2.90~/~21.8  &~2.85~/~18.7  \\ 
		\Xhline{3\arrayrulewidth}
	\end{tabular}
\end{table*}

\subsection{ASV Performance Evaluation} \label{sec5:B}
In this subsection, we compare the proposed TSO-based VACE-WPE front-ends with the single-channel WPE algorithm, which is one of the most commonly adopted front-ends for far-field ASV tasks \cite{drvbf_ffsr18,stc_voices19}.
The ASV performance on the three considered benchmarks is summarized in Table \ref{tab:resnet34_wpe}.
The SNRs and SRMRs of the processed signals are also reported to provide broad estimates of the objective speech quality.
In the table, prefixes ``$\textrm{TSO}_\mathcal{C}\textrm{-}$" and ``$\textrm{DR-TSO}_\mathcal{C}\textrm{-}$" denote the proposed TSO and DR-TSO methods, both of which employ the clean (noise-free) signal as the desired target.
Moreover, prefix ``$\textrm{TSO}_\mathcal{N}\textrm{-}$" denotes a TSO employing a noisy target signal.
As explained in Section \ref{sec3:A}, all $\textrm{TSO}_\mathcal{C}$, $\textrm{DR-TSO}_\mathcal{C}$, and $\textrm{TSO}_\mathcal{N}$ were performed in addition to the Drv-VACE-WPE.
Concurrently, $\textrm{TSO}_\mathcal{C}\textrm{-}$WPE-single was obtained by conducting the proposed TSO in addition to the pretrained neural WPE, with the unfrozen LPSNet parameters.

On the VOiCES benchmark, despite the highest SNR and SRMR measures, the Dns-VACE-WPE is the only candidate inferior to the unprocessed case.
This suggests that direct regression of a clean speech using the VACE-WPE is clearly not desired.
Among the front-ends built without the TSO, the WPE-single and the Drv-VACE-WPE showed overall comparable performance, with the latter exhibiting a considerably higher SRMR scores than the former.
Concurrently, the $\textrm{TSO}_\mathcal{N}\textrm{-}$VACE-WPE could only marginally outperform the WPE-single and the Drv-VACE-WPE.
Interestingly, the SRMR was slightly reduced on both the ``dev" ($2.87 > 2.80$) and ``eval" ($2.73 > 2.63$) sets after conducting the $\textrm{TSO}_\mathcal{N}$.
In contrast, both the $\textrm{TSO}_\mathcal{C}\textrm{-}$VACE-WPE and $\textrm{DR-TSO}_\mathcal{C}\textrm{-}$VACE-WPE achieved significant performance improvements of $18\%$--$31\%$ and $13\%$--$22\%$ relative to the unprocessed case and the WPE-single on the ``eval" dataset, respectively, in terms of both the minDCF and EER.
The improvements achieved on the ``dev" set were also substantial, ranging from $4\%$ to $22\%$.
However, both the SNR and SRMR measures were reasonably smaller than those achieved by the Dns-VACE-WPE, indicating that the degree of denoising and dereverberation desired by the $\textrm{(DR-)TSO}_\mathcal{C}$ differs from the results of the direct regression of clean speech.
Concurrently, compared to the Drv-VACE-WPE and the $\textrm{TSO}_\mathcal{N}\textrm{-}$VACE-WPE, the increments in the SRMR were only marginal, whereas those in the SNR were moderately large.
This implies that the superiority of the $\textrm{(DR-)TSO}_\mathcal{C}\textrm{-}$VACE-WPE is attributed to the denoising capability of the VACE-WPE, but the property should be exploited within a TSO framework to achieve robust ASV performance.
The $\textrm{TSO}_\mathcal{C}\textrm{-}$VACE-WPE moderately outperformed the $\textrm{DR-TSO}_\mathcal{C}\textrm{-}$VACE-WPE in terms of both ASV metrics on the ``eval" partition, likely because the dereverberation capability of the latter was strongly restricted by the DR term described in Eq.\,(\ref{eqn:L_dr}).
Finally, the $\textrm{TSO}_\mathcal{C}\textrm{-}$WPE-single performance was only comparable to that of the WPE-single, suggesting a clear advantage of adopting the VACE-WPE front-end over the single-channel WPE within the proposed TSO framework.

On the SITW benchmark, the ASV performances of all front-ends, except the Dns-VACE-WPE, did not differ significantly from each other.
This was probably because the recording conditions of the SITW and VoxCeleb2 are similar \cite{sitw16,voxceleb2}, and thus, there could have been a slight mismatch between the training and test conditions.
Moreover, the noise and reverberation levels of the SITW datasets were lower than those of the VOiCES datasets, as exhibited by the SNR and SRMR measures.
Nonetheless, only the WPE-single and $\textrm{TSO}_\mathcal{N}\textrm{-}$VACE-WPE consistently improved the unprocessed case on both the ``dev" and ``eval" sets in terms of both ASV metrics.
Concurrently, the $\textrm{TSO}_\mathcal{C}\textrm{-}$VACE-WPE exhibited an overall performance degradation with relative increments of $6\%$--$10\%$ in EER, whereas the $\textrm{DR-TSO}_\mathcal{C}\textrm{-}$VACE-WPE slightly reduced the minDCF of the unprocessed case on both the ``dev" and ``eval" sets as well as the EER on the ``eval" set.
This suggests that without the DR term expressed in Eq.\,(\ref{eqn:L_dr}), the $\textrm{TSO}_\mathcal{C}\textrm{-}$VACE-WPE will produce undesired signal distortions against the input signals with relatively low levels of reverberation, leading to performance degradation.

On the FFSVC benchmark, despite the mismatch between the training datasets for the front-end and DSE model, the three TSO variants of the VACE-WPE\footnote{Our PyTorch codes of VACE-WPE and the pretrained models are available from \texttt{\url{https://github.com/dreadbird06/tso_vace_wpe}}} overall improved the unprocessed case.
In particular, the $\textrm{DR-TSO}_\mathcal{C}\textrm{-}$VACE-WPE consistently slightly outperformed the others, achieving relative improvements of $4.4\%$--$4.5\%$ on the ``dev" set and $1.3\%$--$3.6\%$ on the ``eval" set compared to the unprocessed case.
These smaller improvements, compared to those achieved on the VOiCES datasets, could be attributed to the language mismatch (i.\,e., English vs.\ Chinese Mandarin) as well as the short durations of the utterances being evaluated ($3.1$ s $<$ $15.8$ s).
Interestingly, compared to the Drv-VACE-WPE, the SRMR scores slightly decreased after the $\textrm{(DR-)TSO}_\mathcal{C}$; 
when we checked the individual scores, these decrements in the SRMR originated from relatively clean utterances with high SNRs, and the opposite was caused by relatively noisy ones.
Concurrently, the front-ends built without the TSO, except the WPE-single, generally failed to improve the unprocessed case.

\renewcommand{\arraystretch}{1.05}
\begin{table*}[]
	\caption{
		ASV performance on VOiCES Challenge 2019, SITW core--core, and FFSVC 2020 Task 2 (ResNet-34, fully neural front-ends)
	}
	\label{tab:resnet34_neural}
	
	\centering
	\begin{tabular}{@{}|M{2.7cm}|M{2.3cm}|M{1.5cm}M{1.5cm}|M{1.5cm}M{1.5cm}|M{1.5cm}M{1.5cm}|@{}}
		\Xhline{3\arrayrulewidth}
		\multirow{2}{*}{Method} &
		\multirow{2}{*}{Metric} &
		\multicolumn{2}{c|}{VOiCES 2019 \cite{voices_challenge_2019}} &
		\multicolumn{2}{c|}{SITW \cite{sitw16}} &
		\multicolumn{2}{c|}{FFSVC 2020 \cite{ffsvc20}} \\ \cline{3-8}
		&         & dev   & eval   & dev   & eval  & dev   & eval  \\ 
		\Xhline{2.6\arrayrulewidth}
		\multirow{1}{*}{Drv-CAN}
		& minDCF~/~EER(\%)  &0.141~/~1.23  &0.453~/~6.43  &0.135~/~1.41  &0.160~/~\textbf{1.56}  &0.649~/~5.46  &0.643~/~6.46  \\ 
		\Xhline{1.0\arrayrulewidth}
		\multirow{1}{*}{Drv-DUNet1}
		& minDCF~/~EER(\%)  &0.133~/~1.23  &0.418~/~5.84  &\textbf{0.132}~/~1.38  &0.156~/~\textbf{1.56}  &0.603~/~5.51  &0.625~/~6.24  \\ 
		\Xhline{1.0\arrayrulewidth}
		\multirow{2}{*}{Drv-DUNet2}
		& minDCF~/~EER(\%)  &0.123~/~1.13  &0.403~/~5.71  &0.135~/~\textbf{1.34}  &0.155~/~\textbf{1.53}  &0.595~/~5.09  &0.604~/~6.04  \\ 
		& ~SRMR~/~SNR(dB)   &~2.97~/~14.6  &~2.84~/~10.6  &~2.99~/~20.8  &~3.01~/~20.2  &~2.99~/~14.2  &~2.90~/~12.0  \\ 
		\Xhline{2.6\arrayrulewidth}
		\multirow{1}{*}{DFL-CAN}
		& minDCF~/~EER(\%)  &0.118~/~1.20  &0.443~/~6.61  &0.145~/~1.67  &0.176~/~1.82  &0.590~/~5.25  &0.612~/~6.40  \\ 
		\Xhline{1.0\arrayrulewidth}
		\multirow{1}{*}{DFL-DUNet1}
		& minDCF~/~EER(\%)  &0.152~/~1.54  &0.638~/~9.71  &0.206~/~2.49  &0.278~/~3.08  &0.714~/~7.13  &0.803~/~9.87  \\ 
		\Xhline{1.0\arrayrulewidth}
		\multirow{2}{*}{DFL-DUNet2}
		& minDCF~/~EER(\%)  &0.136~/~1.38  &0.512~/~6.96  &0.188~/~2.41  &0.233~/~2.63  &0.681~/~6.49  &0.753~/~9.25  \\ 
		& ~SRMR~/~SNR(dB)   &~\underline{3.27}~/~\underline{41.8}  &~\underline{3.45}~/~\underline{35.9}  &~\underline{3.28}~/~\underline{32.7}  &~\underline{3.34}~/~\underline{33.8}  &~\underline{3.25}~/~\underline{29.9}  &~\underline{3.26}~/~\underline{25.8}  \\ 
		\Xhline{2.6\arrayrulewidth}
		\multirow{1}{*}{VID-CAN}
		& minDCF~/~EER(\%)  &0.142~/~1.24  &0.344~/~4.78  &0.164~/~1.56  &0.181~/~1.85  &0.659~/~5.40  &0.650~/~6.82  \\ 
		\Xhline{1.0\arrayrulewidth}
		\multirow{1}{*}{VID-DUNet1}
		& minDCF~/~EER(\%)  &0.190~/~1.75  &0.371~/~4.73  &0.165~/~1.63  &0.193~/~2.00  &0.672~/~6.42  &0.669~/~7.02  \\ 
		\Xhline{1.0\arrayrulewidth}
		\multirow{2}{*}{VID-DUNet2}
		& minDCF~/~EER(\%)  &0.154~/~1.59  &0.337~/~4.66  &0.172~/~1.70  &0.183~/~1.85  &0.683~/~5.67  &0.640~/~6.68  \\ 
		& ~SRMR~/~SNR(dB)   &~2.34~/~16.1  &~2.20~/~11.8  &~2.94~/~20.5  &~2.96~/~19.8  &~2.16~/~20.7  &~2.14~/~18.6  \\ 
		\Xhline{2.6\arrayrulewidth}
		\multirow{1}{*}{TSO$_\mathcal{C}$-CAN}
		& minDCF~/~EER(\%)  &0.122~/~1.16  &0.321~/~4.39  &0.147~/~1.56  &0.171~/~1.88  &0.627~/~5.08  &0.633~/~6.65  \\ 
		\Xhline{1.0\arrayrulewidth}
		\multirow{1}{*}{TSO$_\mathcal{C}$-DUNet1}
		& minDCF~/~EER(\%)  &0.116~/~1.13  &0.294~/~3.94  &0.135~/~1.42  &0.159~/~1.77  &\textbf{0.587}~/~4.62  &\textbf{0.591}~/~6.07  \\ 
		\Xhline{1.0\arrayrulewidth}
		\multirow{2}{*}{TSO$_\mathcal{C}$-DUNet2}
		& minDCF~/~EER(\%)  &\textbf{0.111}~/~\textbf{1.09}  &\textbf{0.283}~/~\textbf{3.81}  &0.136~/~1.38  &0.162~/~1.78  &0.616~/~4.68  &0.610~/~6.34  \\ 
		& ~SRMR~/~SNR(dB)   &~2.94~/~25.4  &~2.94~/~20.4  &~3.12~/~27.6  &~3.16~/~26.8  &~2.90~/~22.6  &~2.88~/~20.1  \\ 
		\Xhline{2.6\arrayrulewidth}
		\multirow{1}{*}{DR-TSO$_\mathcal{C}$-CAN}
		& minDCF~/~EER(\%)  &0.119~/~1.12  &0.326~/~4.74  &0.140~/~1.44  &0.158~/~1.64  &0.602~/~4.68  &0.607~/~6.17  \\ 
		\Xhline{1.0\arrayrulewidth}
		\multirow{1}{*}{DR-TSO$_\mathcal{C}$-DUNet1}
		& minDCF~/~EER(\%)  &0.116~/~1.13  &0.295~/~4.00  &0.134~/~1.36  &\textbf{0.152}~/~1.57  &\textbf{0.577}~/~\textbf{4.55}  &\textbf{0.586}~/~\textbf{5.88}  \\ 
		\Xhline{1.0\arrayrulewidth}
		\multirow{2}{*}{DR-TSO$_\mathcal{C}$-DUNet2}
		& minDCF~/~EER(\%)  &\textbf{0.112}~/~\textbf{1.08}  &\textbf{0.283}~/~\textbf{3.85}  &\textbf{0.133}~/~\textbf{1.34}  &\textbf{0.152}~/~\textbf{1.56}  &0.617~/~4.64  &0.604~/~6.22  \\ 
		& ~SRMR~/~SNR(dB)   &~2.93~/~25.1  &~2.92~/~20.0  &~2.99~/~22.7  &~3.02~/~22.0  &~2.89~/~22.0  &~2.87~/~19.5  \\ 
		\Xhline{3\arrayrulewidth}
	\end{tabular}
\end{table*}

\subsection{Comparison With Fully Neural Front-ends} \label{sec5:C}
In this subsection, to further verify the effectiveness of adopting the VACE-WPE as the front-end for far-field ASV tasks, we discuss its comparison with fully neural front-ends.
More specifically, two different DNN models as well as two additional TSO methods were considered.
The two DNN models are as follows:
%
\begin{itemize}
	\item \textit{Context aggregation network (CAN)} \cite{dfl_fe20}: performs 	feature enhancement and consists of multiple cascades of dilated Conv2D, BN \cite{batchnorm}, leaky ReLU \cite{leaky_relu15}, and temporal squeeze-and-excitation \cite{dfl_fe20} operations.
	\item \textit{Dense-U-Net (DUNet)} \cite{dl_target_cancel_wang20}: operates with the RI components of the STFT coefficients and builds a U-Net based on the densely connected Conv2D layers with ELUs \cite{elu_2015} and instance normalization \cite{instance_norm16}; 
	two BLSTM layers are positioned in the bottleneck layer \cite{dl_target_cancel_wang20}.
\end{itemize}
The two TSO methods are described below.
\begin{itemize}
	\item \textit{Deep feature loss (DFL)} \cite{dfl_fe20}: is similar to the proposed TSO framework, except that the loss function is defined in the hidden activation space of the first few layers of the pretrained DSE model.
	The MAE is used instead of the cosine similarity \cite{dfl_fe20}.
	The enhancement DNN model is trained from scratch without pretraining \cite{dfl_fe20}.
	\item \textit{VoiceID (VID) loss} \cite{voiceidloss19}: employs the cross-entropy, defined using the speaker posteriors obtained from the pretrained DSE model, as the loss function.
	The enhancement DNN model is also trained from scratch \cite{voiceidloss19}.
\end{itemize}

The ASV performance regarding the fully neural front-ends is summarized in Table \ref{tab:resnet34_neural}.
Both the CAN and DUNet models were trained using both the DFL and VID loss (prefixes ``DFL-" and ``VID-").
To implement the VID loss, we reinitialized the output layer of the ResNet-34 DSE model and retrained it using the modified LibriSpeech-80h dataset.
Moreover, to extend the proposed TSO and DR-TSO frameworks to build the fully neural front-ends, we first pretrained each model to only perform dereverberation by employing a ``noisy" early arriving target signal (prefix ``Drv-").
In this pretraining stage, the MSE between the output and target features was employed as the loss function for the CAN, whereas the MAE in the RI components of the STFT coefficients as well as that in the linear-scale magnitude spectra were adopted for the DUNet \cite{dl_target_cancel_wang20}.
Subsequently, we performed the TSO and DR-TSO of the pretrained model using the loss functions expressed in Eqs.\,(\ref{eqn:L_tso}) and (\ref{eqn:L_dr_tso}), respectively (prefixes ``$\textrm{TSO}_\mathcal{C}\textrm{-}$" and ``$\textrm{DR-TSO}_\mathcal{C}\textrm{-}$").
Two different DUNet models, DUNet1 and DUNet2, were implemented.
The CAN and DUNet1 operated with frame and hop sizes of $25$\,ms and $10$\,ms, respectively, to match those of the DSE model, but the DUNet2 with $64$\,ms and $16$\,ms, respectively, to implement the drop-in replacement of the VACE-WPE in Fig.\,\ref{fig:tso} with a fully neural front-end.
The DUNet1 did not apply the inverse STFT for speaker embedding extraction; 
rather, MFBEs were directly computed on the enhanced spectra.
The CAN had $11.29$ million parameters, with the channel size fixed as $60$, and the DUNet1 and DUNet2 had $16.13$ and $20.55$ million parameters, respectively.

First, on the VOiCES ``eval" dataset, all ``DFL-" variants failed to improve the unprocessed case, with the DFL-DUNet1 being particularly inferior to the others.
In contrast, 
all ``$\textrm{TSO}_\mathcal{C}\textrm{-}$" and ``$\textrm{DR-TSO}_\mathcal{C}\textrm{-}$" variants successfully improved the unprocessed case.
Thus, it can be inferred that pretraining the model to achieve dereverberation, but not denoising, plays an important role by inductively biasing the model to preserve the noise components, potentially avoiding the undesired signal distortions produced by the fully neural speech denoisers.
To the best of our best knowledge, no previous studies have proposed explicitly pretraining a neural network front-end for only dereverberation, without denoising, prior to conducting a TSO for robust far-field ASV.
Nonetheless, unlike the WPE-single or the Drv-VACE-WPE, the fully neural models trained for dereverberation without a TSO (i.\,e., the prefix ``Drv-") could not improve the unprocessed case.
Interestingly, despite random initialization, the ``VID-" variants did not generally undergo performance degradation on the ``eval" set.
This is probably because unlike the DFL that promotes to learn a denoising mechanism between a pair of corrupted and clean signals, the VID loss does not necessarily require an explicit denoising operation of a front-end, but rather focuses on minimizing the cross-entropy loss of a pretrained speaker classifier.
Thus, with the DSE model itself being reasonably robust to (acoustic) noise, the VID loss could have implicitly avoided exccessive reduction of noise, and hence deterioration of ASV performance.
Indeed, the SRMR and SNR measures of the VID-DUNet2 were substantially lower than those of other denoising candidates, whereas the DFL-DUNet2 exhibited the highest scores.
%
Nonetheless, on the ``dev" partition, only the $\textrm{TSO}_\mathcal{C}\textrm{-}$DUNet2 and $\textrm{DR-TSO}_\mathcal{C}\textrm{-}$DUNet2 could marginally decrease the minDCF of the unprocessed case.
Concurrently, within the proposed $\textrm{TSO}_\mathcal{C}$ and $\textrm{DR-TSO}_\mathcal{C}$ frameworks, the DUNet models marginally outperformed the CAN models on both the ``dev" and ``eval" datasets, with the DUNet2 models being slightly superior to the DUNet1 models.
Contrary to the $\textrm{TSO}_\mathcal{C}\textrm{-}$VACE-WPE and the $\textrm{DR-TSO}_\mathcal{C}\textrm{-}$VACE-WPE performance differences observed in Table \ref{tab:resnet34_wpe}, no significant gap was found between the ``$\textrm{TSO}_\mathcal{C}\textrm{-}$" and ``$\textrm{DR-TSO}_\mathcal{C}\textrm{-}$" variants of the DUNet models.
However, both were generally inferior to their VACE-WPE counterparts.
%

Second, on the SITW benchmark, all ``DFL-" and ``VID-" variants significantly degraded the unprocessed signal performance, whereas the ``$\textrm{TSO}_\mathcal{C}\textrm{-}$" and ``$\textrm{DR-TSO}_\mathcal{C}\textrm{-}$" variants exhibited comparable performance.
This indicates that, to reduce the undesired signal distortions caused by the fully neural front-ends, a TSO must be preceded by a pretraining of a front-end in a noise-preserving direction.
Concurrently, there was a noticeable performance difference between the ``$\textrm{TSO}_\mathcal{C}\textrm{-}$" and ``$\textrm{DR-TSO}_\mathcal{C}\textrm{-}$" variants, with the minDCFs and EERs of the latter respectively being $0.74\%$--$7.6\%$ and $2.9\%$--$12.8\%$ relatively lower compared to the former.
This suggests that the strategy of preserving hardly reverberant input signals during the TSO, as suggested by Eq.\,(\ref{eqn:L_dr}), is also effective for building a fully neural front-end.
Among the ``$\textrm{DR-TSO}_\mathcal{C}\textrm{-}$" variants, the $\textrm{DR-TSO}_\mathcal{C}\textrm{-}$DUNet2 slightly outperformed the others and was favorably comparable to the $\textrm{DR-TSO}_\mathcal{C}\textrm{-}$VACE-WPE.
Finally, all the results demonstrate the difficulty of establishing a fully neural front-end for both far-field and ``in-the-wild" ASV purposes, particularly when there is a mismatch in training datasets for the front-end and the DSE model.

Contrary to that observed in Table \ref{tab:resnet34_wpe}, all fully neural variants led to performance degradation of the unprocessed case on the FFSVC benchmark, suggesting that the VACE-WPE could be more robust or insensitive to the condition mismatches than the fully neural front-ends.
This may be attributed to the operation of the former being only partially dependent on the neural network but strictly restricted by the WPE algorithm.
In comparison, the latter produces the output signal solely by the cascade of the neural network layers, all of which were trained using a completely data-driven approach.
Among all fully neural candidates, the $\textrm{DR-TSO}_\mathcal{C}\textrm{-}$DUNet1 exhibited the lowest minDCF and EER than the others.

\begin{figure}[t]
	\centering
	\includegraphics[width=\linewidth]{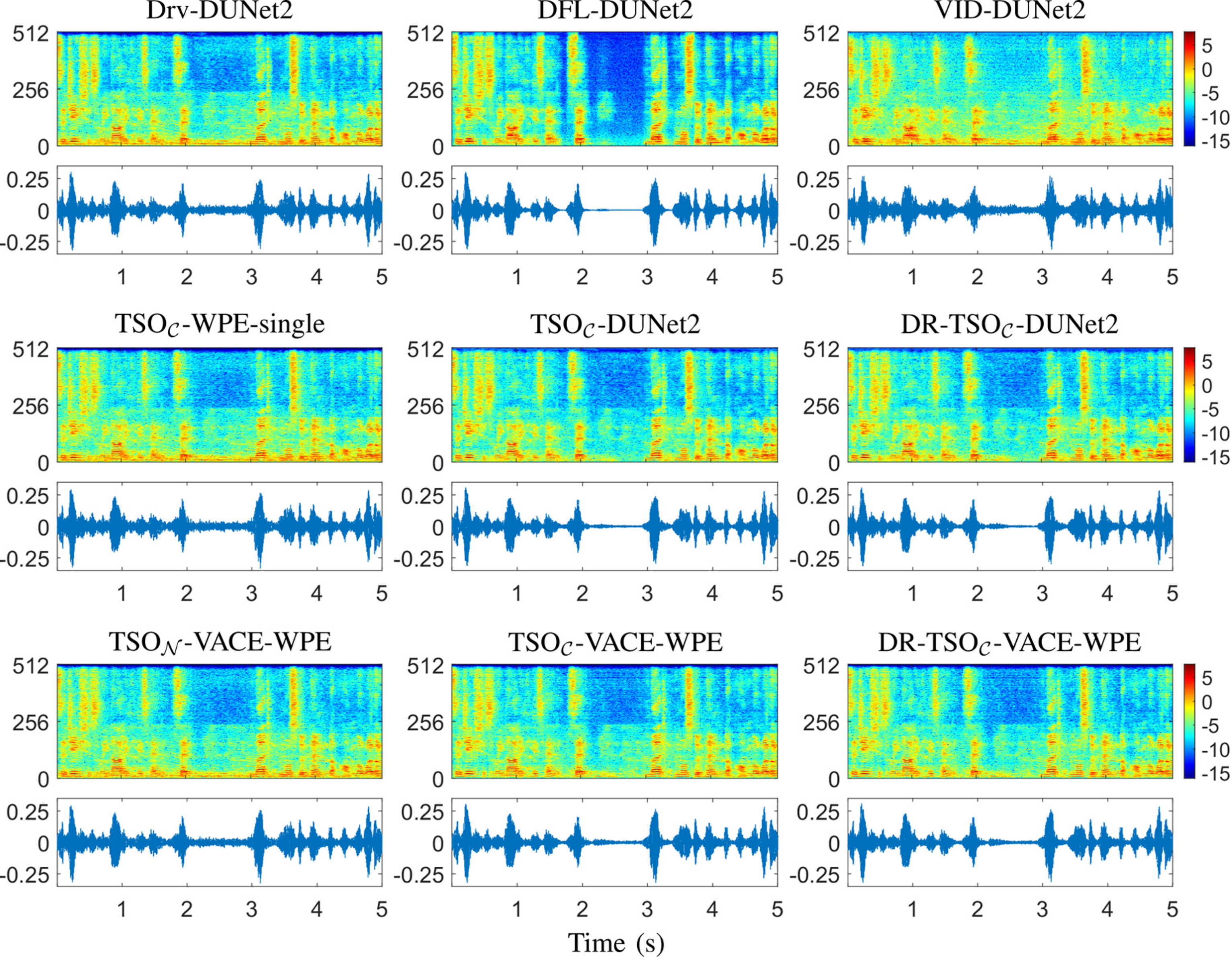}
	\caption{
		Comparison of the processed signals obtained from the variants of the WPE-single, VACE-WPE, and DUNet front-ends.
	}
	\label{fig:output_signals}
\end{figure}

Fig.\,\ref{fig:output_signals} shows the output signals of the various front-ends for the same speech excerpt presented in Fig.\,\ref{fig:comp_dns_drv}.
The figures show that the $\textrm{(DR-)TSO}_\mathcal{C}$ variants definitively remove the noise components from the input signal to some extent, except for the $\textrm{TSO}_\mathcal{C}\textrm{-}$WPE-single.
Although the neural WPE was subjected to the $\textrm{TSO}_\mathcal{C}$, the $\textrm{TSO}_\mathcal{C}\textrm{-}$WPE-single could not explicitly remove the noise.
Concurrently, the DFL-DUNet2 suppressed the noise more aggressively than the others, whereas the VID-DUNet2 exhibited substantial amount of residual noise.
This supports our analysis on the denoising mechanism of the DFL and VID loss provided with the ASV results on the VOiCES datasets.

\renewcommand{\arraystretch}{1.07}
\begin{table*}[]
	\caption{
		ASV performance on VOiCES Challenge 2019, SITW core--core, and FFSVC 2020 Task 2 (ECAPA-TDNN, WPE-based front-ends)
	}
	\label{tab:ecapa_wpe}
	
	\centering
	\begin{tabular}{@{}|M{2.7cm}|M{2.3cm}|M{1.5cm}M{1.5cm}|M{1.5cm}M{1.5cm}|M{1.5cm}M{1.5cm}|@{}}
		\Xhline{3\arrayrulewidth}
		\multirow{2}{*}{Method} &
		\multirow{2}{*}{Metric} &
		\multicolumn{2}{c|}{VOiCES 2019 \cite{voices_challenge_2019}} &
		\multicolumn{2}{c|}{SITW \cite{sitw16}} &
		\multicolumn{2}{c|}{FFSVC 2020 \cite{ffsvc20}} \\ \cline{3-8}
		&         & dev   & eval   & dev   & eval  & dev   & eval  \\ 
		\Xhline{2.6\arrayrulewidth}
		\multirow{1}{*}{Unprocessed}
		& minDCF~/~EER(\%)  &0.121~/~1.17  &0.311~/~5.02  &0.148~/~1.42  &0.148~/~1.45  &0.648~/~5.59  &0.629~/~6.30  \\ 
		\Xhline{2.6\arrayrulewidth}
		\multirow{1}{*}{WPE-single}
		& minDCF~/~EER(\%)  &0.108~/~1.09  &0.280~/~4.78  &0.145~/~1.44  &0.143~/~1.66  &0.636~/~5.41  &0.626~/~6.27  \\ 
		\Xhline{1\arrayrulewidth}
		\multirow{1}{*}{TSO$_\mathcal{C}$-WPE-single}
		& minDCF~/~EER(\%)  &0.109~/~1.08  &0.280~/~4.78  &0.144~/~1.45  &0.144~/~1.66  &0.636~/~5.54  &0.623~/~6.31  \\ 
		\Xhline{2.6\arrayrulewidth}
		\multirow{1}{*}{Dns-VACE-WPE}
		& minDCF~/~EER(\%)  &0.132~/~1.28  &0.361~/~5.22  &0.212~/~2.08  &0.229~/~2.41  &0.754~/~7.85  &0.781~/~8.88  \\ 
		\Xhline{1\arrayrulewidth}
		\multirow{1}{*}{Drv-VACE-WPE}
		& minDCF~/~EER(\%)  &0.112~/~1.08  &0.286~/~4.80  &0.145~/~1.43  &\textbf{0.143}~/~\textbf{1.46}  &0.663~/~5.56  &0.638~/~6.51  \\ 
		\Xhline{2.6\arrayrulewidth}
		\multirow{1}{*}{TSO$_\mathcal{N}$-VACE-WPE}
		& minDCF~/~EER(\%)  &0.108~/~\textbf{1.07}  &0.268~/~4.66  &0.145~/~1.42  &0.146~/~1.61  &0.637~/~5.24  &0.619~/~\textbf{6.14}  \\ 
		\Xhline{1\arrayrulewidth}
		\multirow{1}{*}{TSO$_\mathcal{C}$-VACE-WPE}
		& minDCF~/~EER(\%)  &\textbf{0.106}~/~1.11  &\textbf{0.246}~/~\textbf{4.20}  &\textbf{0.144}~/~\textbf{1.42}  &0.148~/~1.72  &\textbf{0.628}~/~5.17  &0.617~/~6.16  \\ 
		\Xhline{1\arrayrulewidth}
		\multirow{1}{*}{DR-TSO$_\mathcal{C}$-VACE-WPE}
		& minDCF~/~EER(\%)  &\textbf{0.106}~/~1.11  &\textbf{0.249}~/~\textbf{4.22}  &0.147~/~1.45  &0.144~/~1.50  &0.631~/~\textbf{5.12}  &\textbf{0.612}~/~6.21  \\ 
		\Xhline{3\arrayrulewidth}
	\end{tabular}
\end{table*}


\renewcommand{\arraystretch}{1.07}
\begin{table*}[]
	\caption{
		ASV performance on VOiCES Challenge 2019, SITW core--core, and FFSVC 2020 Task 2 (ECAPA-TDNN, fully neural front-ends)
	}
	\label{tab:ecapa_neural}
	
	\centering
	\begin{tabular}{@{}|M{2.7cm}|M{2.3cm}|M{1.5cm}M{1.5cm}|M{1.5cm}M{1.5cm}|M{1.5cm}M{1.5cm}|@{}}
		\Xhline{3\arrayrulewidth}
		\multirow{2}{*}{Method} &
		\multirow{2}{*}{Metric} &
		\multicolumn{2}{c|}{VOiCES 2019 \cite{voices_challenge_2019}} &
		\multicolumn{2}{c|}{SITW \cite{sitw16}} &
		\multicolumn{2}{c|}{FFSVC 2020 \cite{ffsvc20}} \\ \cline{3-8}
		&         & dev   & eval   & dev   & eval  & dev   & eval  \\ 
		\Xhline{2.6\arrayrulewidth}
		\multirow{1}{*}{Drv-DUNet1}
		& minDCF~/~EER(\%)  &0.137~/~1.30  &0.316~/~5.09  &\textbf{0.146}~/~1.43  &\textbf{0.145}~/~\textbf{1.48}  &0.695~/~6.27  &0.661~/~\textbf{6.62}  \\ 
		\Xhline{1\arrayrulewidth}
		\multirow{1}{*}{Drv-DUNet2}
		& minDCF~/~EER(\%)  &0.135~/~1.23  &0.315~/~5.08  &\textbf{0.147}~/~1.45  &0.147~/~1.62  &0.691~/~6.02  &0.655~/~\textbf{6.62}  \\ 
		\Xhline{2.6\arrayrulewidth}
		\multirow{1}{*}{DFL-DUNet1}
		& minDCF~/~EER(\%)  &0.188~/~1.83  &0.433~/~5.80  &0.243~/~2.71  &0.274~/~2.92  &0.848~/~9.67  &~0.899~/~12.56  \\ 
		\Xhline{1\arrayrulewidth}
		\multirow{1}{*}{DFL-DUNet2}
		& minDCF~/~EER(\%)  &0.133~/~1.33  &0.377~/~5.73  &0.201~/~2.23  &0.215~/~2.30  &0.792~/~8.29  &~0.845~/~11.31  \\ 
		\Xhline{2.6\arrayrulewidth}
		\multirow{1}{*}{VID-DUNet1}
		& minDCF~/~EER(\%)  &0.178~/~1.76  &0.346~/~5.00  &0.190~/~1.78  &0.192~/~1.86  &0.713~/~7.05  &0.729~/~7.71  \\ 
		\Xhline{1\arrayrulewidth}
		\multirow{1}{*}{VID-DUNet2}
		& minDCF~/~EER(\%)  &0.164~/~1.77  &0.326~/~4.80  &0.190~/~1.85  &0.186~/~1.87  &0.698~/~6.64  &0.706~/~7.08  \\ 
		\Xhline{2.6\arrayrulewidth}
		\multirow{1}{*}{TSO$_\mathcal{C}$-DUNet1}
		& minDCF~/~EER(\%)  &0.125~/~1.21  &0.271~/~4.43  &0.154~/~1.46  &0.153~/~1.51  &0.661~/~\textbf{5.64}  &\textbf{0.644}~/~6.76  \\ 
		\Xhline{1\arrayrulewidth}
		\multirow{1}{*}{TSO$_\mathcal{C}$-DUNet2}
		& minDCF~/~EER(\%)  &\textbf{0.122}~/~\textbf{1.19}  &\textbf{0.266}~/~\textbf{4.20}  &0.157~/~\textbf{1.35}  &0.159~/~1.61  &\textbf{0.651}~/~6.04  &0.670~/~7.17  \\ 
		\Xhline{2.6\arrayrulewidth}
		\multirow{1}{*}{DR-TSO$_\mathcal{C}$-DUNet1}
		& minDCF~/~EER(\%)  &0.125~/~1.20  &0.284~/~4.46  &0.151~/~1.50  &0.148~/~\textbf{1.46}  &0.658~/~\textbf{5.50}  &\textbf{0.641}~/~\textbf{6.57}  \\ 
		\Xhline{1\arrayrulewidth}
		\multirow{1}{*}{DR-TSO$_\mathcal{C}$-DUNet2}
		& minDCF~/~EER(\%)  &\textbf{0.121}~/~\textbf{1.18}  &\textbf{0.268}~/~\textbf{4.23}  &0.152~/~1.43  &\textbf{0.146}~/~\textbf{1.48}  &\textbf{0.649}~/~5.94  &0.661~/~6.95  \\ 
		\Xhline{3\arrayrulewidth}
	\end{tabular}
\end{table*}

\subsection{Generalization to Unseen DSE Model} \label{sec5:D}
In practical scenarios, building a front-end without access to the DSE model of the target ASV system is a general requirement.
To simulate a scenario in which the DSE models employed for the ASV and TSO do not match and further verify the proposed method in that scenario, we used the pretrained DSE model described in \cite{speechbrain21} to implement the ASV system, whereas the ResNet-34 DSE model was employed to only perform the TSO.
More specifically, the DSE model of \cite{speechbrain21} was built with the recently proposed emphasized channel attention, propagation, and aggregation (ECAPA) time-delay neural network (TDNN) \cite{ecapa_tdnn20}.
The following are some of the details of the ECAPA-TDNN model:
80D log-MFBEs extracted with window, hop, and FFT sizes of $25$\,ms, $10$\,ms, and $400$, respectively, were the input features.
A mini-batch of 3 s time-domain speech segments was augmented on-the-fly using real RIRs \cite{kaldi_rirs}, MUSAN \cite{musan_corpus} noises, speed perturbation \cite{speed_perturb15}, and time-domain random masking \cite{ecapa_tdnn20}.
Additive angular margin softmax \cite{arcface19} was employed with a margin of $0.2$ and a scale of $30$.
The Adam optimizer \cite{adam_optimizer} was used to train the model using a triangular cyclic learning rate scheduler \cite{cyclic_lr17}.
Because the model configurations of the ECAPA-TDNN and the ResNet-34 DSE model differ significantly, we believe that the former can be considered unseen DSE model.
Note that the ECAPA-TDNN was additionally fine-tuned using the FFSVC 2020 ``train" partition, as described in Section \ref{sec4:C:1}, prior to measuring the ASV performance on the FFSVC 2020 benchmark.
The ECAPA-TDNN had $20.77$ million parameters.

Table \ref{tab:ecapa_wpe} summarizes the ASV performance of the unseen ECAPA-TDNN model equipped with WPE-based front-ends.
Trends similar to those observed in Table \ref{tab:resnet34_wpe} from the ResNet-34 DSE model were found on all three benchmarks.
Both the $\textrm{TSO}_\mathcal{C}\textrm{-}$VACE-WPE and $\textrm{DR-TSO}_\mathcal{C}\textrm{-}$VACE-WPE generally outperformed other methods on the VOiCES datasets, particularly by a large margin on the ``eval" partition.
Unlike the results shown in Table \ref{tab:resnet34_wpe}, they presented comparable performance in terms of both the minDCF and EER.
However, on the SITW benchmark, a noticeable performance gap was found between them on the ``eval" partition, particularly in terms of the EER ($1.72>1.50$).
Concurrently, on the FFSVC benchmark, all TSO-based VACE-WPE variants systematically outperformed the unprocessed case in terms of both ASV metrics, although the improvements were still marginal.
Comparing the overall performance, the ECAPA-TDNN model was marginally superior to the ResNet-34 model on the VOiCES and SITW datasets, but it was inferior on the FFSVC benchmark with reference to the primary metric.

Table \ref{tab:ecapa_neural} tabulates the results obtained using the ECAPA-TDNN model and fully neural front-ends; 
the CAN variants were excluded because they can only enhance the 64D log-MFBEs.
To deal with the mismatch in the FFT sizes of the front-ends and the ECAPA-TDNN DSE model ($512\!\neq\!400$), the output STFT coefficients of the DUNet1 variants were converted back to the time-domain waveforms for speaker embedding extraction.
Similar to the trends observed based in Tables \ref{tab:resnet34_wpe} and \ref{tab:resnet34_neural}, the $\textrm{TSO}_\mathcal{C}\textrm{-}$VACE-WPE and $\textrm{DR-TSO}_\mathcal{C}\textrm{-}$VACE-WPE (in Table \ref{tab:ecapa_wpe}) moderately outperformed their DUNet counterparts (in Table \ref{tab:ecapa_neural}) on the VOiCES benchmark.
%
Moreover, on the SITW benchmark, the former group marginally outperformed the latter group in terms of the minDCF, but was slightly inferior with regard to the EER.
However, on the FFSVC datasets, all TSO-based DUNet front-ends failed to successfully enhance the raw signal, similar to the trend observed in Table \ref{tab:resnet34_neural} from the (seen) ResNet-34 DSE model.

\begin{figure*}
	\centering
	\subfloat[ResNet-34, VOiCES Challenge 2019 \cite{voices_challenge_2019}]{ {\includegraphics[width=2.3in]{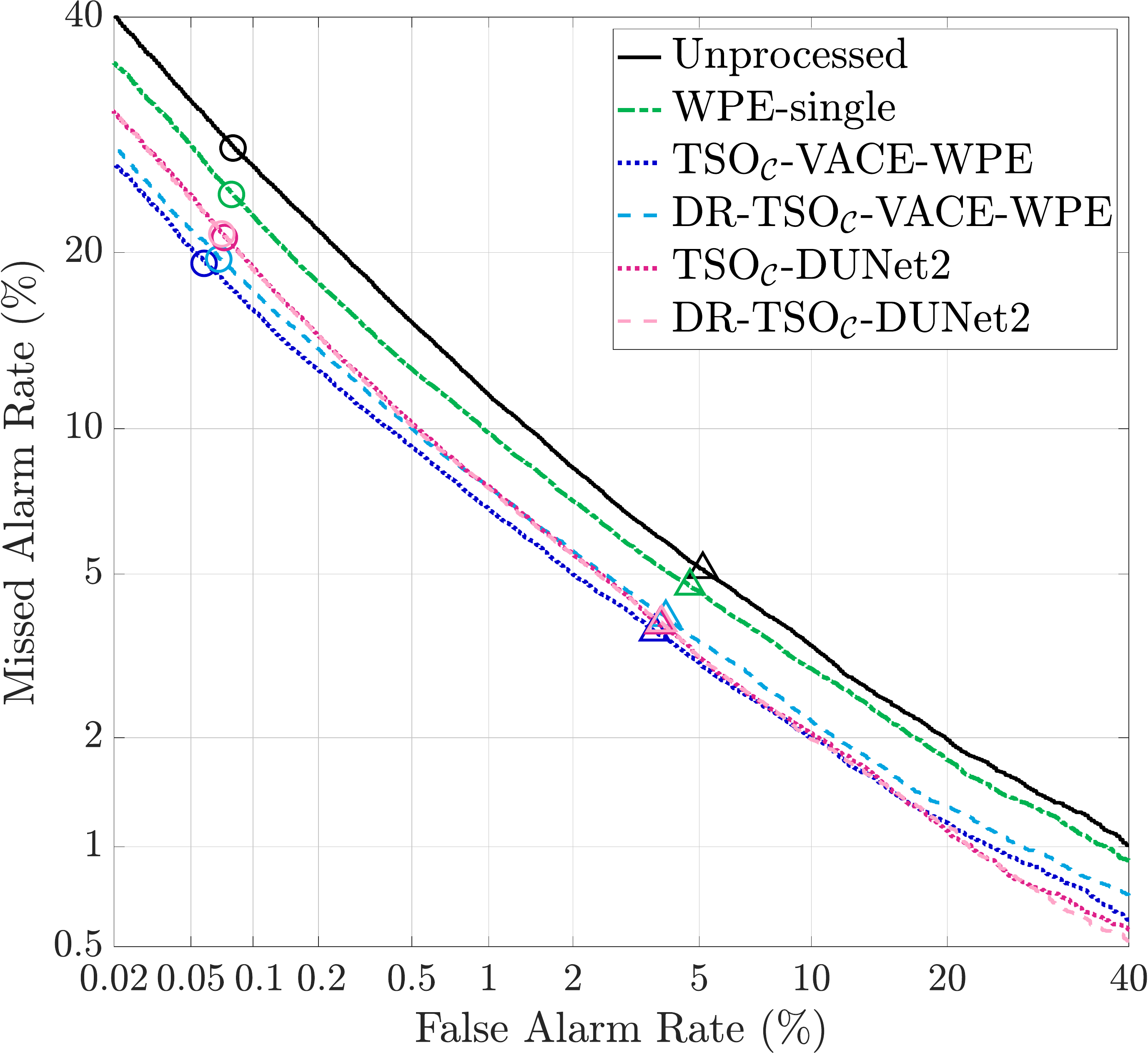}} }
	\subfloat[ResNet-34, SITW \cite{sitw16}]{ {\includegraphics[width=2.3in]{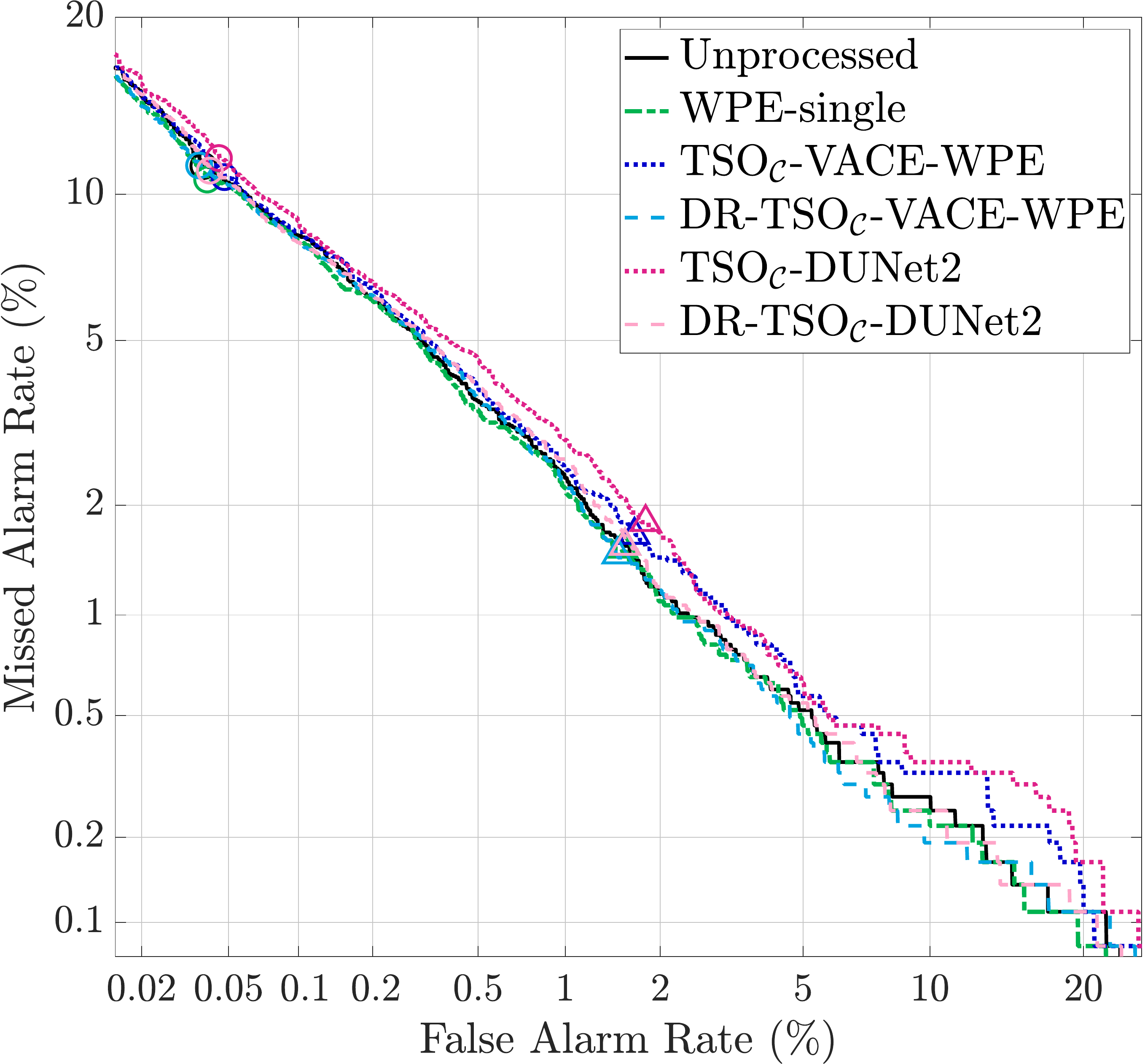}} }
	\subfloat[ResNet-34, FFSVC 2020 \cite{ffsvc20}]{ {\includegraphics[width=2.3in]{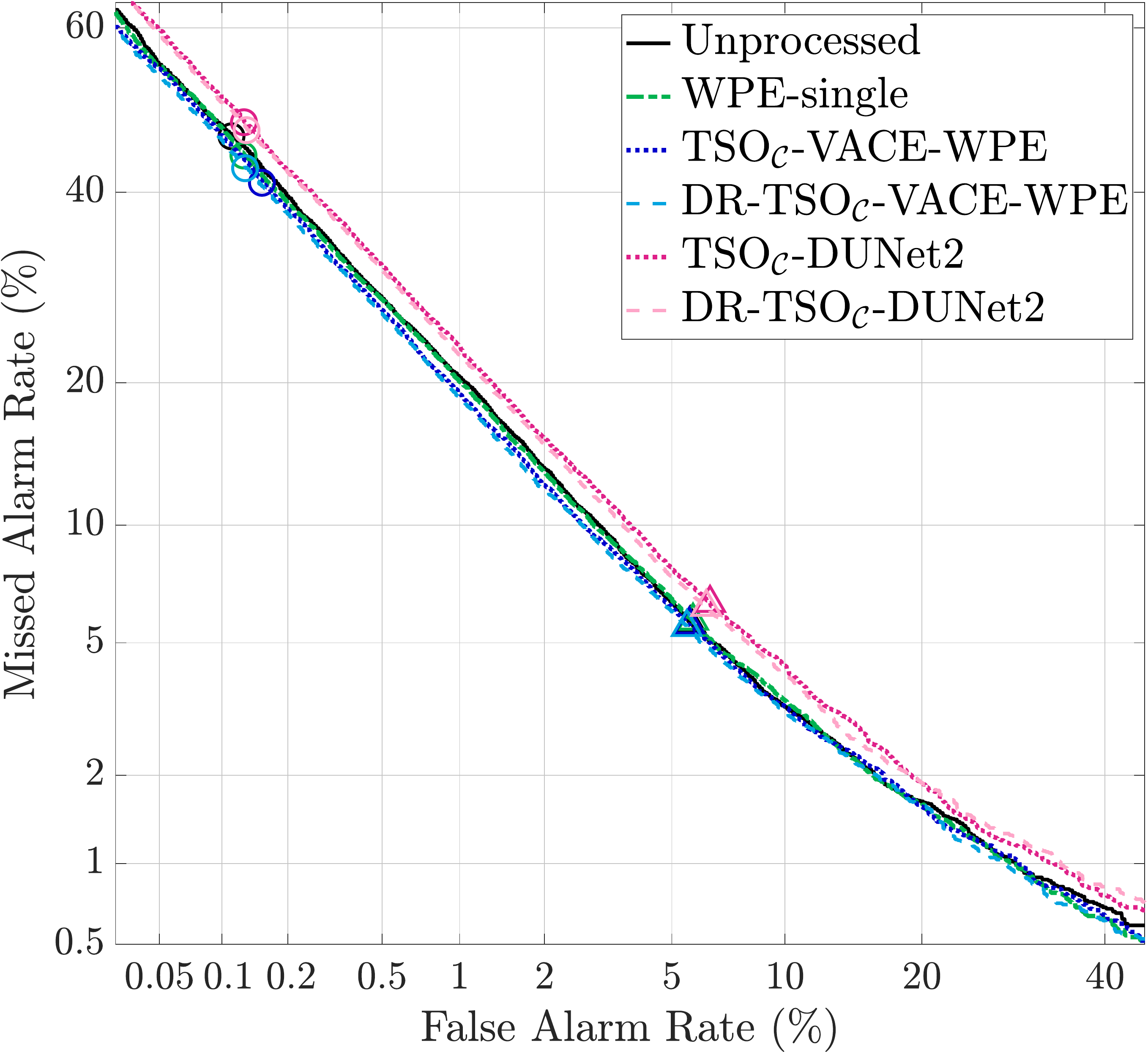}} }
	\hfill
	\subfloat[ECAPA-TDNN, VOiCES Challenge 2019 \cite{voices_challenge_2019}]{ {\includegraphics[width=2.3in]{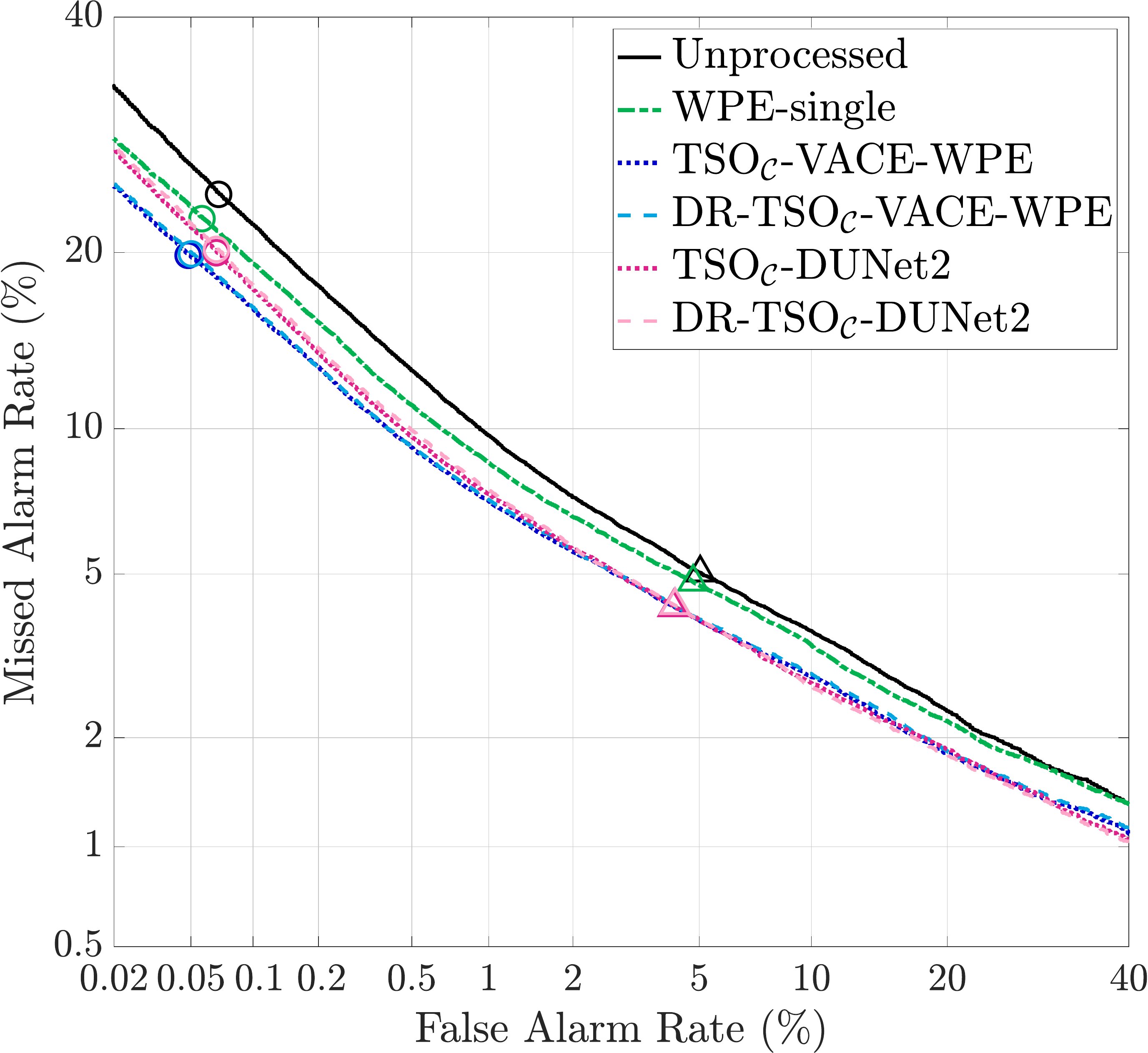}} }
	\subfloat[ECAPA-TDNN, SITW \cite{sitw16}]{ {\includegraphics[width=2.3in]{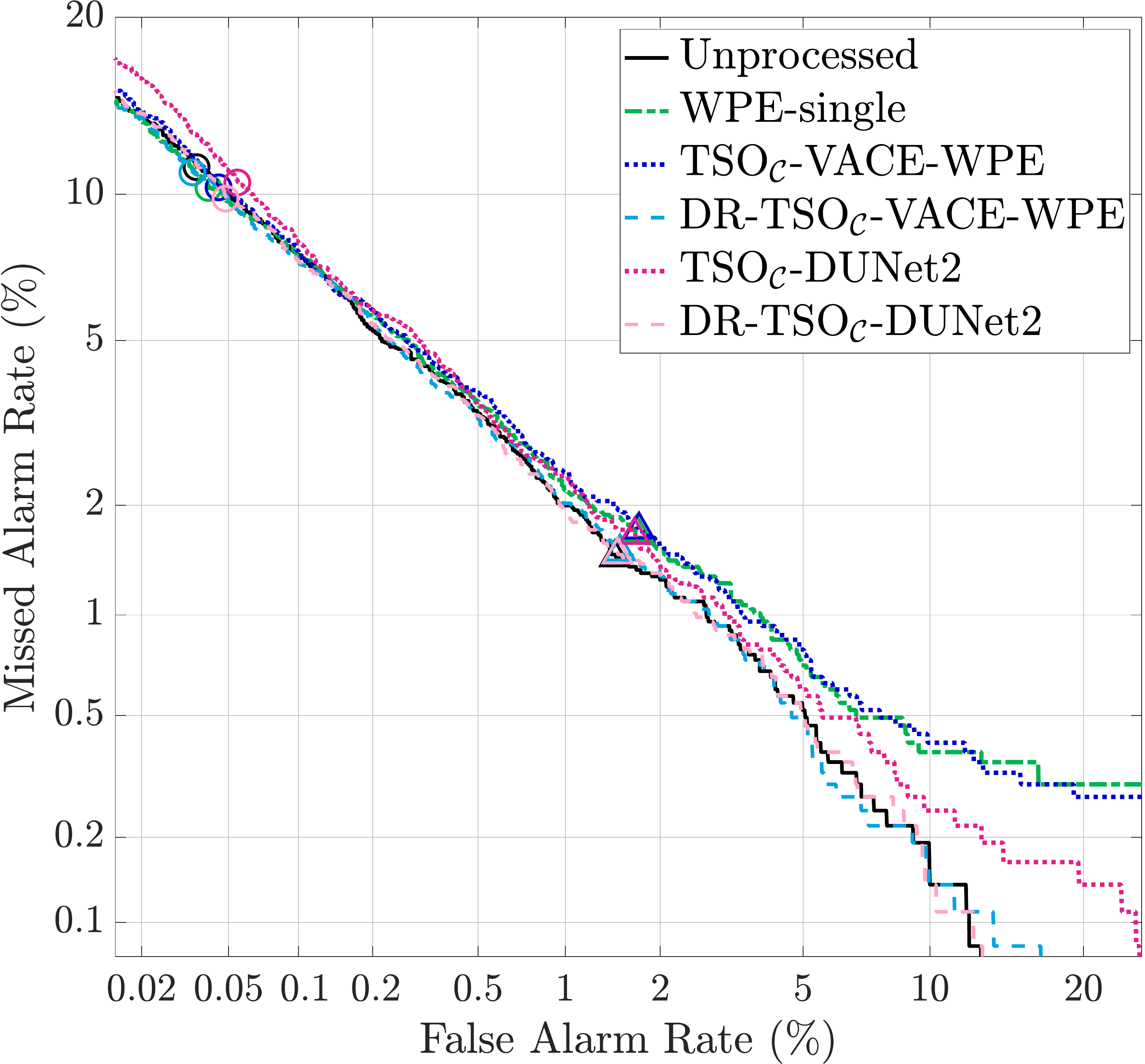}} }
	\subfloat[ECAPA-TDNN, FFSVC 2020 \cite{ffsvc20}]{ {\includegraphics[width=2.3in]{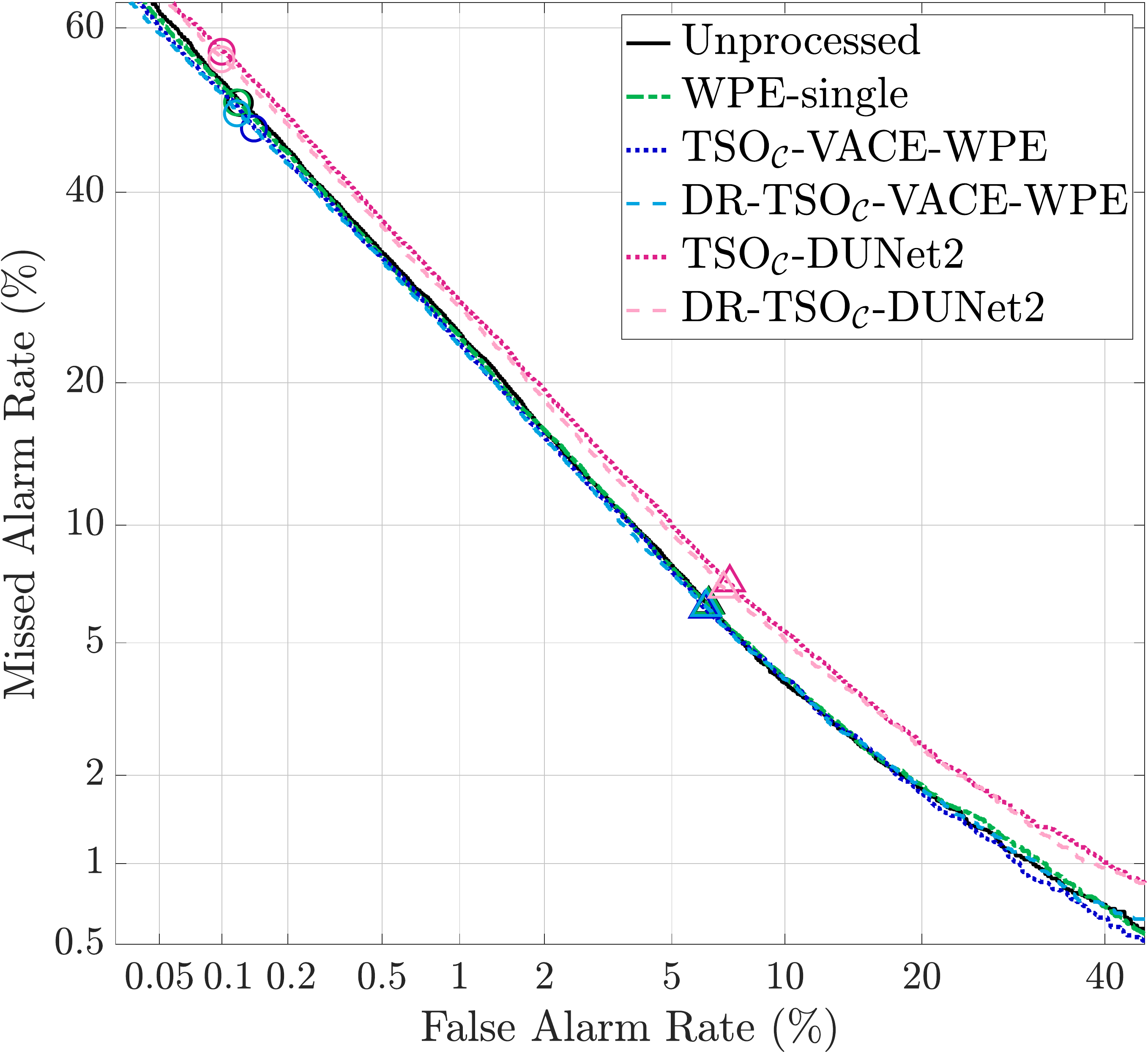}} }
	\caption{
		DET curves of ResNet-34- and ECAPA-TDNN-based ASV systems equipped with different front-ends.
		Operating points for minDCF and EER are represented by the symbols ``$\mathbf{\bigcirc}$" and ``$\bigtriangleup$", respectively.
	}
	\label{fig:det_curves}
\end{figure*}

\subsection{Detection Error Tradeoff (DET) Analysis} \label{sec5:E}
In this subsection, we analyze the DET curves of different ASV systems.
DET curves present a wide range of operating points of ASV systems with regard to missed and false alarm rates in addition to those represented by the minDCF and the EER.
Fig.\,\ref{fig:det_curves} shows the DET curves of the ASV systems built with the ResNet-34 and the ECAPA-TDNN DSE models.
Only the ``$\textrm{TSO}_\mathcal{C}\textrm{-}$" and ``$\textrm{DR-TSO}_\mathcal{C}\textrm{-}$" variants of the VACE-WPE and the DUNet2 front-ends were compared because they presented overall best performances but exhibited comparable performance with each other; 
the WPE-single was also included as a baseline algorithm.
%
Because the number of nontarget trials was significantly greater than that of target trials (see Table \ref{tab:db_statistics}), the minDCF measures with equal cost parameters (i.\,e., $C_\textrm{miss}=C_\textrm{fa}=1$) were obtained from the operating points exhibiting relatively high missed alarm rates.

First, on the VOiCES benchmark, the ResNet-34-based ASV systems and the ECAPA-TDNN-based ASV systems exhibited noticeably different DETs.
Specifically, the former was generally superior to the latter in the low missed alarm region and inferior in the low false alarm region.
%
Interestingly, a similar tradeoff was also observed between the VACE-WPE and the DUNet2 variants for both groups of ASV systems.
Second, on the SITW benchmark, the ``$\textrm{DR-TSO}_\mathcal{C}\textrm{-}$" variants presented overall superior performance across different operating points compared to the ``$\textrm{TSO}_\mathcal{C}\textrm{-}$" variants, particularly in the low missed alarm region.
Finally, on the FFSVC benchmark, the DETs of the considered systems were mostly consistent with each other, with the DUNet2 variants exhibiting overall inferior performance compared to the others.

\section{Conclusions} \label{sec6}
In this study, to deal with the challenge of single-channel far-field ASV in noisy reverberant environments, we established a novel front-end design based on the VACE-WPE and the TSO method employing a pretrained DSE model.
Moreover, based on the dereverberation regime of the WPE algorithm, we proposed a DR-TSO, which further enforces the front-end to preserve the input signals with low levels of reverberation.
Both the proposed TSO and DR-TSO were supported by the denoising capability of the VACE-WPE, a property newly identified in this study by conducting denoising experiments of the VACE-WPE.
The ASV experiments were mainly divided into three parts: 
\romannumeral1) comparison of the proposed front-ends and other WPE variants; 
\romannumeral2) comparison to fully neural front-ends, with an extension of the proposed TSO framework to them; 
and 
\romannumeral3) exploring a mismatch scenario of the DSE models employed for the TSO and ASV.
All experiments were conducted on two far-field ASV benchmarks, with English and Chinese Mandarin, respectively, and on one in-the-wild ASV benchmark to further examine the generalizability of the front-ends to more natural, unconstrained ASV scenarios.
The experimental results suggest that the proposed TSO methods are effective for constructing both the VACE-WPE and fully neural front-ends, with the former being superior to the latter within the proposed TSO framework, particularly in mismatched conditions.
In addition, the proposed DR-TSO was effective for building a front-end robust to both far-field and in-the-wild conditions.
We believe that the proposed $\textrm{DR-TSO}_\mathcal{C}\textrm{-}$VACE-WPE may be a promising direction for developing a unified signal enhancement front-end to support a blind ASV system in various scenarios.
%




 
\bibliographystyle{IEEEtran}
\bibliography{tso_vwpe}

\end{document}